\documentclass[a4paper,11pt]{article}
\pdfoutput=1 % if your are submitting a pdflatex (i.e. if you have
\usepackage{jcappub} % for details on the use of the package, please
\usepackage{caption}
\usepackage{subcaption}   
\usepackage{hyperref}

\usepackage[T1]{fontenc} % if needed
\usepackage[section]{placeins}

\hypersetup{
    colorlinks=true,
    linkcolor=blue,
    filecolor=magenta,      
    urlcolor=cyan,
    pdftitle={Overleaf Example},
    pdfpagemode=FullScreen,
    }

\title{The quest for CMB signatures of  Conformal Cyclic Cosmology}

\author[a]{Eve Bodnia,}
\author[b]{Vlad Isenbaev,}
\author[a]{Kellan Colburn,}
\author[a]{Joe Swearngin,}
\author[a, c]{Dirk Bouwmeester}

\affiliation[a]{Department of Physics, University of California, Santa Barbara, California 93106, USA}
\affiliation[b]{Nuro, Inc., 1300 Terra Bella Ave., Mountain View, USA}
\affiliation[c]{Huygens-Kamerlingh Onnes Laboratorium, Leiden University, 2333 CA, Leiden, Netherlands}

\emailAdd{ebodnia@ucsb.edu}
\abstract{Circles of low-variance and Hawking points in the Cosmic Microwave Background (CMB), resulting from black hole mergers and black hole evaporation, respectively, in a previous cycle of the universe, have been predicted as possible evidence for the Conformal Cyclic Cosmology model (CCC) introduced by R. Penrose. We present a high-resolution search for such low-variance circles  in the Planck and WMAP CMB data, and introduce HawkingNet, our machine learning open-source software based on a ResNet18 algorithm, to search for Hawking points in the CMB.
We find that spots consisting of a few unusually bright (high-temperature) or dark (low-temperature) pixels, erroneously lead to regions with many low-variance circles, and consequently sets of near-concentric low-variance circles,  when applying the search criteria used in previous work \cite{CCC_Penrose_Gurzadyan}. After removing those spots from the data, no statistically significant low-variance circles can be found. 
Concerning Hawking points, also no statistically significant evidence is found when using a Gaussian temperature amplitude model over $\sim 1^{\circ}$ opening angle and after accounting for spots of unusual brightness.
That the unusual spots in the data are themselves  remnants of Hawking points is not supported by low-variance and/or low-temperature circles around them. 
The absence of such statistically-significant distinct features in the currently available CMB data does not disprove the CCC model, but implies that higher resolution CMB data and/or refined CCC based predictions are needed to pursue the search for CCC signatures further.

}

\begin{document}


\maketitle
\flushbottom

\section{Introduction}

The Conformal Cyclic Cosmology model (CCC) introduced by R. Penrose \cite{Penrose_CCC_book,CCC_Intro}, and inspired by work of P. Tod \cite{tod, Tod2013TheEO}, led to the qualitative prediction of concentric low-variance circular rings in the cosmic microwave background (CMB) resulting from super-massive black-hole mergers in the previous aeon - a former version of our current Universe \cite{CCC_Penrose_Gurzadyan}. Furthermore, Hawking evaporation of black holes, the expected final fate of isolated super-massive black holes in the previous aeon, are predicted to give rise to bright, approximately Gaussian shaped, spots in the CMB, referred to as Hawking points. Qualitative arguments predict that Hawing points are surrounded by concentric low-variance circular rings that are somewhat lower in temperature than the average CMB temperature, and that concentric low-variance rings without a central Hawking point should be somewhat higher in temperature \cite{CCC_Penrose_Gurzadyan}.
 
%First indications of the existence of the low variance circles in the CMB data have been previously reported by several groups \cite{CCC_Penrose_Gurzadyan, paper1, paper2}.
V. Gurzadyan and R. Penrose (GP) reported that they found numerous families of concentric low-variance rings in the Wilkinson Microwave Anisotropy Probe (WMAP)- the release of 7 years CMB data \cite{CCC_Penrose_Gurzadyan, WMAP_7_years}. They concluded, by comparing with numerical simulations, that these circles are statistically relevant structures in the CMB data, rather than resulting from Gaussian random noise, thus providing evidence of conformal cyclic cosmology \cite{CCC_Penrose_Gurzadyan_comments}. They compared the numbers of the ring families with the corresponding numbers of the elliptically distorted rings in the CMB, which is referred in their text as the "sky-twist" procedure. In the CMB data, GP examined 10,885 circular patches in the sky of equal number of pixels, systematically covering the sky but excluding the region of our galactic disk. 

In the GP method, the temperature variance, i.e. the standard deviation $\sigma$ in the temperature, was calculated over circular rings of annular width 0.5 degree and of angular radii in the range 2.5-16 degrees in steps of 0.5 degree. If among those rings around a given centre, the temperature variance was at least 15 $\mu K$ below the average variance for the rings between 2.5 and 16 degrees angular radii for that specific centre, the ring was considered "low variance". Many instances of 2, 3 and also of 4 concentric circles where identified in Ref.~\cite{CCC_Penrose_Gurzadyan}, seemingly providing support to the CCC hypothesis. 

We point out that the results in Refs.~\cite{CCC_Penrose_Gurzadyan,CCC_Penrose_Gurzadyan_comments} were obtained by using the WMAP data with galactic plane removal only. No mask was used to separating the CMB data from foreground emission and instrumental effects, although it is stated in Ref.~\cite{CCC_Penrose_Gurzadyan} without further elaboration that using the KQ85 mask (Ref.~\cite{KQ85}) makes no significant difference to the main conclusions. 

%A possible motivation for this choice is the expectation that the detection of CMB features of interest, ring-like structures of significant size (with angular radii in the range 2.5-16 degrees), would hardly be influenced by a foreground mask. One might further argue that any additional data filtering (removing foreground emission and instrumental effects) might reduce the visibility of the signatures of CCC. 

%%%Unusually bright or dark pixels are known to reside within the CMB and are referred to as CMB anomalies. Such anomalies might be the result of the extensive data processing, or of inaccuracies in the measuring system, or perhaps represent real, but yet unknown, astronomical phenomena.

% (this section needs lots of rephrasing. Better summary found in https://arxiv.org/pdf/1508.05158.pdf)

The occurrence of low-variance circles in the same WMAP data was reexamined in 2010, Ref.~\cite{paper1} and in 2011,  Ref.~\cite{paper2}. The authors did confirmed the presence of regions of concentric circles in the CMB data, and found that the WMAP signal is indeed dominated by a cosmological signal rather than by instrumental noise. However, they claimed that such variations are the result of usual CMB anisotropies and therefore statistically expected. We note that the simulation approach by these groups is very different from the one taken by GP. GP's way of simulating data is constructing the map of the Gaussian sky injected with the WMAP’s noise, in contrast to the groups of Refs.~\cite{paper1, paper2} who simulate the data using the CMB power spectrum plus the WMAP noise for the Lambda cold dark matter model ($\Lambda CDM$).
Gurzadyan \emph{et al.} argued that the treatment of the CMB signal as random Gaussian noise is not fully justified, based on a Kolmogorov analyzis of the degree of randomness in CMB, see Ref.~ \cite{invalid_gaussian}. The search for low variance circles was furthermore repeated in 2015, Ref.~\cite{planck_ccc}, on the WMAP data set and on the newer Planck data. The authors again confirmed the existence of special directions with near-concentric circles of low-variance claimed by Ref.~\cite{CCC_Penrose_Gurzadyan}; however, they also considered those consistent with the predictions of the standard cosmological model. For their analysis, the authors used the same parameter set from Ref.~\cite{CCC_Penrose_Gurzadyan} on the WMAP W-band smoothed with a 20 arcminute full-width at half-maximum (FWHM) Gaussian beam data set and the newer Planck COMMANDER data. The COMMANDER data were downgraded from 2048 pixels to 512 and also smoothed with a 20 arcminute FWHM Gaussian beam to save computation time. The simulations were created by the Planck best-fit $\Lambda CDM$ angular power spectra generated for 2048 pixels and also downgraded to 512 pixels and smoothed out with the same Gaussian profile, see Ref.~\cite{Planck_power_spec_2018}. 
%The authors discovered that the distribution of low variance rings on the hemispheres is prevalent in the COMMANDER data, but concluded the asymmetry to be statistically insignificant, given that nearly $25\%$ of their simulations produce a distribution of low variance directions with a higher dipole asymmetry than seen in the data. The origin of the CMB dipole asymmetry evidence based on WMAP and Planck data has been questioned in Ref.~\cite{CMB_string_2}, and possible explanations are ranging from inhomogeneous noise distributions at the considered scales to the exotic dark matter theories and cosmic strings, see Refs.~\cite{CMB_string, CMB_string_2, CMB_string_3, CMB_string_4}. 

In the first part of this article we will show that the procedure followed by GP to identify (concentric) circles of low variance is prone to incorrect identification of such (concentric) circles.   
Because references~\cite{paper1,paper2,planck_ccc} used variations of the GP method they seem to have erroneously confirmed the presence of the (concentric) circles, but with different statistical significance because they did use proper foreground masks and performed different simulations for their comparison with the real CMB data analysis.  
To understand how false (concentric) circles of low-variance appear from the GP method, note the decisive role of the average of the variance of the rings with annular width 0.5 degree and of angular radii in the range 2.5-16 degrees in steps of 0.5 degree, which we will refer to as the baseline variance, in determining which circles are considered to be of low variance. If for some reason this baseline variance is unexpectedly high, what can happen when there is an unusually high or low temperature spot of a few pixels in the data in the region from 2.5 to 16 degrees, there will be low-variance circles identified that would not be there without this bright or dark spot. The erroneously identified circles will come in clusters, making it likely to observe families of near-concentric low-variance circles. Therefore, as we will show, regions of large numbers of (concentric) circles of low variance are traced back to just a few unusually bright of dark pixels.

Even after applying a foreground mask, it turns out that a few of the bright/dark spots remain and cause the incorrect appearance of regions of (near-concentric) circles of low variance. That is most likely the reason why GP could claim that using a KQ85 mask did not make a significant difference to their main conclusions, and also why the other groups confirmed the presence of the (concentric) circles of low variance.

The second part of this paper concerns the CCC prediction of Hawking points. Hawking points have been studied in Refs.~\cite{Melissa_HP, Hawking_Penrose, HP_Jow}. According to CCC, in the previous Aeon (a cycle of the Universe), (almost) all of the mass eventually falls into supermassive black holes, which then evaporate via Hawking radiation. At its late life stage, the Aeon undergoes enormous conformal compression, forcing the remnant Hawking radiation to become Hawking points at the crossover surface. The Hawking points are predicted to show up in the CMB sky in the form of a spatially Gaussian signal spread over a small data patch (of the order of $\sim 1^{\circ}$ opening angle) with an amplitude of the order of 100 mK, see Ref.~\cite{Hawking_Penrose}. Reference \cite{Hawking_Penrose} claims strong observational evidence of numerous previously unobserved anomalous circular spots of significantly raised temperature in the CMB. Reference \cite{Melissa_HP} introduces a new search method for Hawking points and the findings are reported to be inconclusive and in need of further investigation. Reference \cite{HP_Jow} claims that an excess in Hawking points is detected in Planck satellite maps at
only an 87 $\%$ confidence level (i.e., little more than $1\sigma$), instead of the 99.98$\%$ confidence level attributed to Ref.~\cite{Hawking_Penrose}, and therefore 
no statistically significant evidence for the presence of Hawking points in the CMB can be concluded.

In summary, there is significant controversy about the claims in support of the CCC model, and we noted a potential pitfall for statistical analyses of CCC signatures set by spots in the WMAP data of a few unusual bright or dark pixels. 

In this article we take advantage of a supercomputer and machine learning approaches to firstly try to reproduce the analysis in Ref.~\cite{CCC_Penrose_Gurzadyan} on WMAP data, without a corresponding foreground map, and the analysis in Ref.~\cite{planck_ccc}, but with 500,000 centers in the sky, instead of 10,885 centers, and without smoothing the data. 
We find that a few (of order 10) spots of unusual brightness in the WMAP data play a decisive role in the data analysis. 
We repeat the search for low-variance circles after removing the small collection of spots of unusual brightness. We furthermore apply the appropriate mask to separate the CMB data from foreground emission and instrumental effects, and compare the results.
Secondly, we search for the Hawking points on high-resolution Planck data sets using different power spectra and a novel ResNet-based machine learning (ML) tool (Refs.~\cite{ML_basis_2, ML_basis}) on over 50 million centers in the sky. %and 1000 runs per simulation.  
Our findings do not support the previous claims of statistically significant sets of low-variance circles and Hawking points. %However, we do observe from our Hawking point study that the CMB temperature fluctuations are locally (over an $\sim 1^{\circ}$ opening angle) smoother than expected from a best-fit $\Lambda CDM$ and COM PowerSpect CMB TT-full power spectra Gaussian noise model. For larger regions of $\sim 2.5-16^{\circ}$ opening angle, as explored for the search for concentric circles of low-variance, a best-fit $\Lambda CDM$ and COM PowerSpect CMB TT-full power spectra Gaussian noise model does provide reasonable results. 

\section{The Gurzadyan Penrose method with high resolution}
We first investigate the search method of Gurzadyan and Penrose (GP) \cite{CCC_Penrose_Gurzadyan}. For this method it is essential to establish an appropriate "baseline" temperature variance for a certain region in the sky around which the variance of ring-like structures is to be determined. As mentioned above, the average (baseline) temperature variance was obtained by averaging the variance for the rings (with a width of 0.5 degree angular radius) between 2.5 and 16 degrees angular radii around a specific centre. We performed the baseline calculations and searched for the low variance circles across both hemispheres using 500,000 centers in both  WMAP and Planck COMMANDER data. Specifically, we used the native healpix nsideparameter 512 for WMAP (7 year W band temperature map) and 2048 for Planck COMMANDER data. We first used the galactic plane only (GPO) mask, in order to be consistent with the data analyses of Ref.~\cite{CCC_Penrose_Gurzadyan}, and next apply a standard foreground mask.
We did not perform any smoothing of the data.

\begin{figure}[h]
    \centering 
    \begin{subfigure}[c]{0.49\textwidth}
         \centering
         \includegraphics[width=\textwidth]{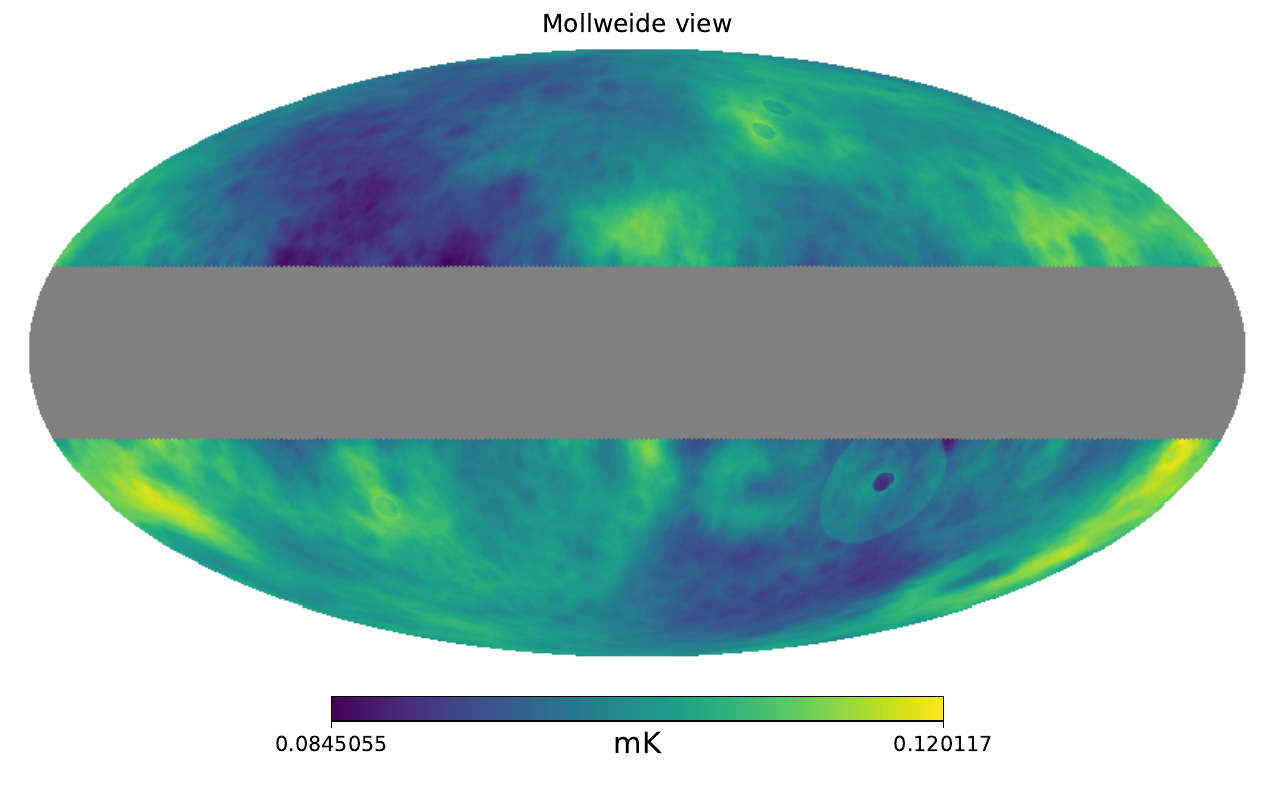}
        %  \caption{$y=5/x$}
        %  \label{fig:five over x}
     \end{subfigure}
        % \caption{Three simple graphs}
        %\label{fig:three graphs}
     \begin{subfigure}[c]{0.49\textwidth}
         \centering
         \includegraphics[width=\textwidth]{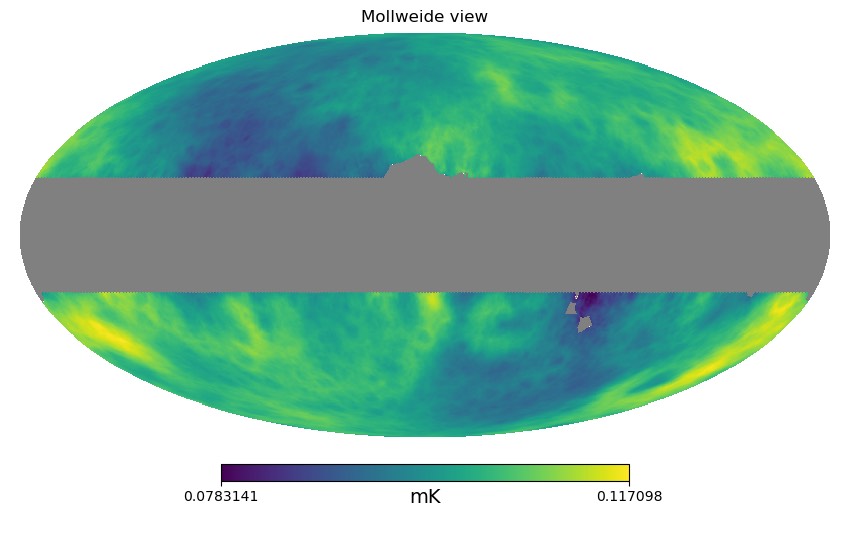}
        %  \caption{$y=5/x$}
        %  \label{fig:five over x}
     \end{subfigure}
     
        \caption{The baseline calculation for the Gurzadyan and Penrose search method on the WMAP data \cite{CCC_Penrose_Gurzadyan}. 
        Left: with the galactic plane only (GPO) mask. Right with the KQ85 mask ("wmap-temperature-kq85-analysis-mask-r9-9yr-v5.fits").   The baseline temperature variance was obtained by averaging the variance for rings (with a width of 0.5 degree angular radius) between 2.5 and 16 degrees angular radii around each specific centre point. }
    \label{fig:compareWMAP}
\end{figure}

\begin{figure}[h]
    \centering 
    \begin{subfigure}[c]{0.49\textwidth}
         \centering
         \includegraphics[width=\textwidth]{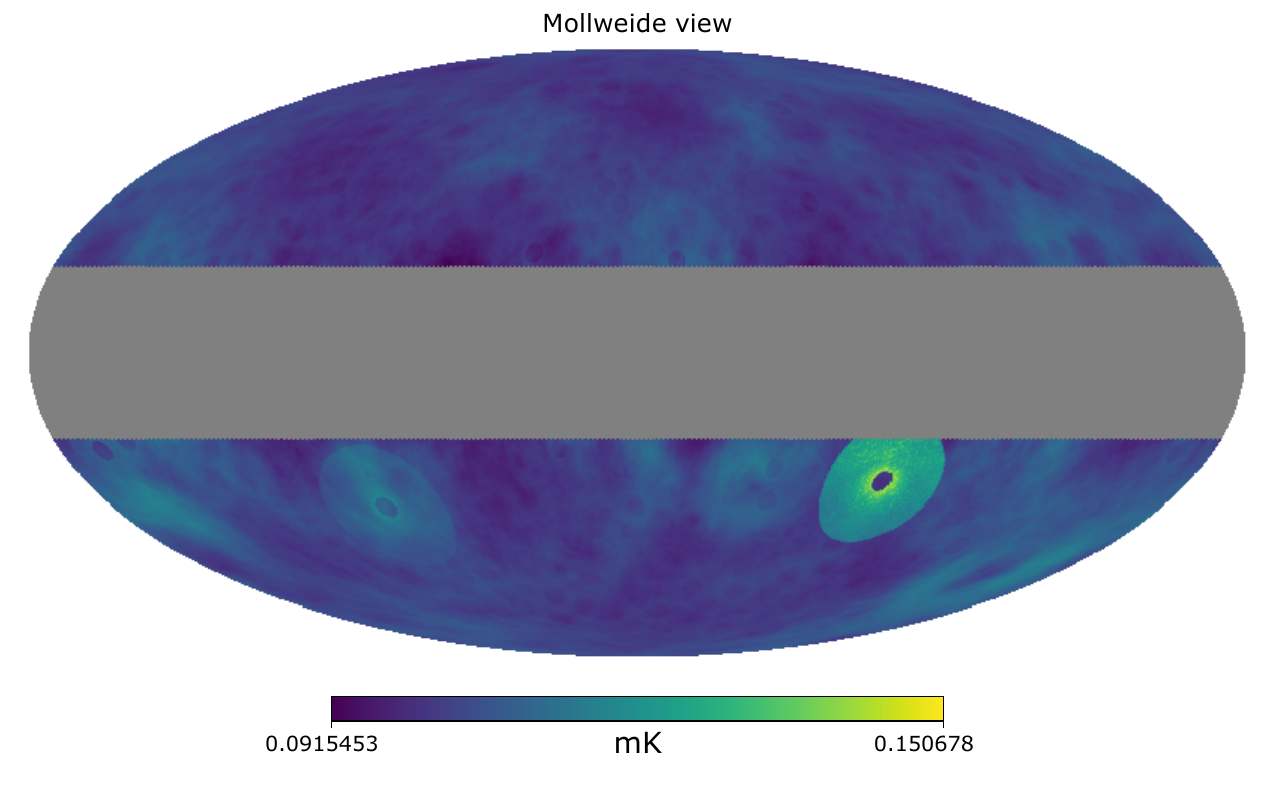}
        %  \caption{$y=5/x$}
        %  \label{fig:five over x}
     \end{subfigure}
        % \caption{Three simple graphs}
        %\label{fig:three graphs}
     \begin{subfigure}[c]{0.49\textwidth}
         \centering
         \includegraphics[width=\textwidth]{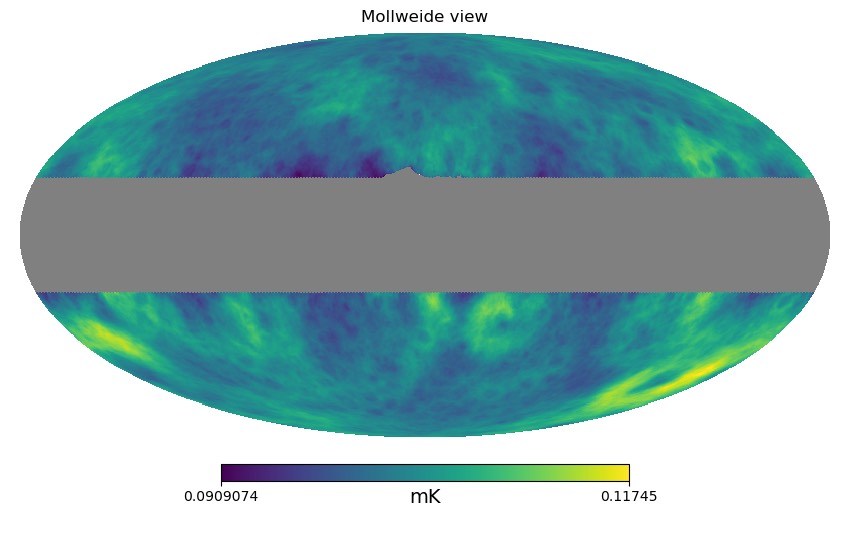}
        %  \caption{$y=5/x$}
        %  \label{fig:five over x}
     \end{subfigure}
     
        \caption{The baseline calculation for the Gurzadyan and Penrose search method on Planck COMMANDER data. Left GPO and right with foreground mask ("COM-Mask-CMB-common-Mask-Int-2048-R3.00.fits").The baseline temperature variance was obtained by averaging the variance for rings (with a width of 0.5 degree angular radius) between 2.5 and 16 degree angular radii around each specific centre point. }
        \label{fig:compareCommander}
\end{figure}

Figure ~\ref{fig:compareWMAP} presents the baseline variance underlying the  GP analysis, of the WMAP data \cite{CCC_Penrose_Gurzadyan}; Left with the galactic plane only (GPO) mask, Right with the KQ85 mask ("wmap-temperature-kq85-analysis-mask-r9-9yr-v5.fits"). 
Figure~\ref{fig:compareCommander} shows the baseline variance in Planck COMMANDER data; Left GPO, Right with foreground mask ("COM-Mask-CMB-common-Mask-Int-2048-R3.00.fits"). Both figures show pronounced higher baseline variance regions with sharp circular boundaries, especially in the panels on the left (without foreground mask). Note that the scale bars indicate changes in the baseline variance by up to $\sim 35 \mu K$ in Fig.~\ref{fig:compareWMAP} left and right, and up to $\sim 60 \mu K$ in Fig.~\ref{fig:compareCommander} left, and $\sim 22 \mu K$ in Fig.~\ref{fig:compareCommander} rigth.

%\begin{figure}[hbt!]
 % \centering
  %\includegraphics[scale=0.7]{pics/PG_WMAP_baseline_25-16.pdf}
  %  \caption{The baseline calculation for the Gurzadyan and Penrose search method on the WMAP data \cite{CCC_Penrose_Gurzadyan}. The baseline temperature variance was obtained by averaging the variance for rings (with a width of 0.5 degree angular radius) between 2.5 and 16 degrees angular radii around each specific centre point.}
  %\label{fig:WMAP_baseline}
  %\end{figure}
%\begin{figure}[hbt!]
%\centering
% \includegraphics[scale=0.7]{pics/our_baseline_mollweide_PlANCK_25-16.pdf}
%\caption{The baseline calculation for the Gurzadyan and Penrose search method on Planck COMMANDER data. The baseline temperature variance was obtained by averaging the variance for rings (with a width of 0.5 degree angular radius) between 2.5 and 16 degree angular radii around each specific centre point.}
  %\label{fig:our_Planck_COMMANDER_baseline}
%\end{figure}

%\begin{figure}[hbt!]
%  \centering
%  \includegraphics[scale=0.7]{pics/PG_WMAP_baseline.pdf}
%    \caption{The baseline calculation for the Gurzadyan and Penrose search method on the WMAP data \cite{CCC_Penrose_Gurzadyan}. The baseline temperature variance was obtained by averaging the variance for rings (with a width of 0.5 degree angular radius) between 16 and 32 degrees angular radii around each specific centre point.}
%  \label{fig:WMAP_baseline}
%  \end{figure}

%\begin{figure}[hbt!]
%\centering
% \includegraphics[scale=0.7]{pics/our_baseline_mollweide_PlANCK.pdf}
%\caption{The baseline calculation for the Gurzadyan and Penrose search method on Planck COMMANDER data. The baseline temperature variance was obtained by averaging the variance for rings (with a width of 0.5 degree angular radius) between 16 and 32 angular radii around each specific centre point.}
% \label{fig:our_Planck_COMMANDER_baseline}
%\end{figure}

\begin{figure}[hbt!]
  \centering
 \includegraphics[scale=0.6]{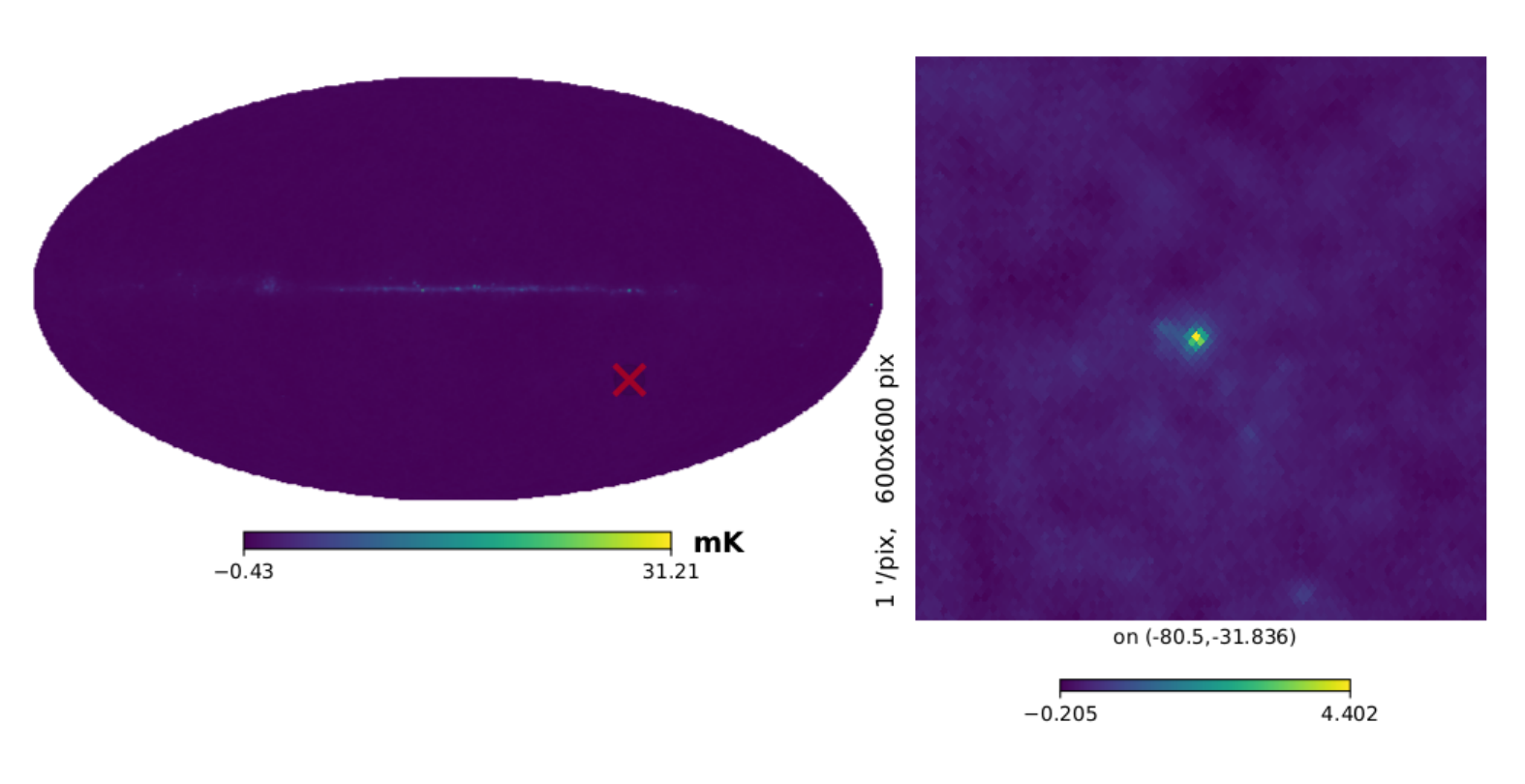}
    \caption{The most pronounced bright spot in the WMAP data, left (and also in the Plank COMMANDER data, left), that significantly raises the baseline variance in temperature (as defined by the GP method) in the surrounding region. Comparing to the mean CMB temperature of approximately 2.725K, the temperature of this spot is 2.7294K.}
  \label{fig:WMAP_outlier}
\end{figure}

The origin of those regions turns out to be tiny regions of a few pixels with a significantly higher or lower temperature, at the center of each ring-shaped structure, combined with the GP method of determining the baseline. Figure \ref{fig:WMAP_outlier} shows, for example,  the spot that gives rise to the most pronounced ring feature of relatively high baseline variance, but this spot is removed by the foreground masks. It is this tiny spot that is entirely responsible for the incorrect identification of a high density of concentric circles of low variance in this region in Ref.\cite{CCC_Penrose_Gurzadyan}

\begin{figure}[hbt!]
  \centering
 \includegraphics[scale=0.5]{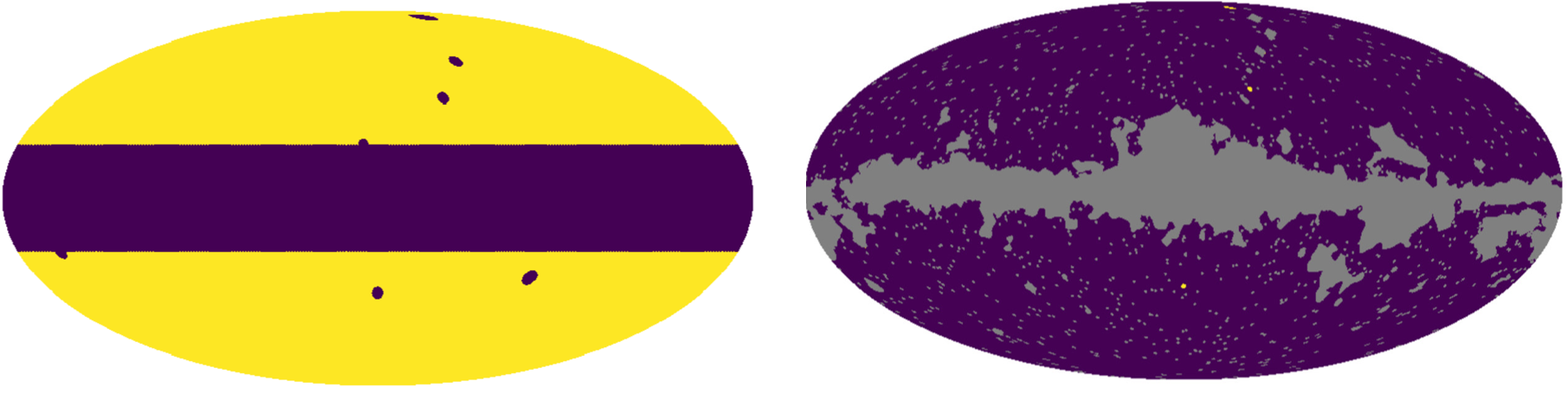}
    \caption{Left: The "baseline filter" which contains the Galactic Plane Only (GPO) mask and in addition removes the dominant spots that give rise to the ring patterns in the baseline map as shown in Fig.~\ref{fig:compareWMAP}, left. Right: After applying the KQ85 mask, still several spots of unusual brightness remain, the yellow points in the Right panel.}
  \label{fig:WMAP_ourmask}
\end{figure}

Figure \ref{fig:WMAP_ourmask} Left shows the locations of the dominant spots that are identified in Fig.~\ref{fig:compareWMAP} Left, as the centers of the ring patterns. We will refer to  this map as the "baseline filter".
Figure \ref{fig:WMAP_ourmask} Right shows the locations of the remaining dominant spots (in yellow), after using the KQ85 mask, that are responsible for some of the ring patterns in Fig.~\ref{fig:compareWMAP} Right. Those remaining spots have likely contributed to, what we argue, is the "erroneous" confirmation of the concentric circles of low variance in Refs.~\cite{paper1,paper2,planck_ccc}.

The consequence of the raised baseline regions is that one will find in those specific regions many low-variance (lower by $15 \mu K$ or more, compared to the baseline) circles that are in fact not real low-variance circles. Those erroneously identified rings will be clustered around the spots with unusual brightness, and therefore have a high probability of generating families of nearly concentric circle. It is very unfortunate that this feature precisely coincides with a qualitative prediction of the CCC model.

To explicitly demonstrate that a few spots of unusual brightness account for many incorrectly identified sets of concentric circles, and to address the question whether or not concentric circles remain after we filter out the few spots of unusual brightness, we proceed with the GP method.

Figure \ref{fig:n-circlesWMAP} (a)-(e) shows our data on finding with the GP method (a) single circles of low variance in the WMAP (7 year W band temperature map, native healpix nsideparameter 512)  CMB data (with radii between 2.5 and 16 degrees angular radii), and (b) 2, (c) 3, (d) 4, and (e) 5 concentric circles of low variance around a specific center point (for 500.000 centers uniformly distributed over the sky, with the exclusion of the galactic plane). 

The purple lines in Fig.~\ref{fig:n-circlesWMAP} (a)-(e) show the data in the case that the spots of unusual brightness are still in the CMB data (and the GPO mask is applied). This is the situation explored by GP \cite{CCC_Penrose_Gurzadyan}. The blue lines show the results when we apply our "baseline filter" (Fig.~\ref{fig:WMAP_ourmask}, Left), which removes the galactic plane and the spots of unusual brightness that are causing the rings in the baseline calculations shown in Fig.~\ref{fig:compareWMAP} Left. The red lines show the data after the foreground KQ85 mask has been applied and some spots of unusual brightness (shown Fig.~\ref{fig:WMAP_ourmask}, Right) are still present.
We see that at the level of individual circles, Fig.~\ref{fig:n-circlesWMAP} (a), the lines almost coincide. When looking at the case for two concentric circles, Fig.~\ref{fig:n-circlesWMAP} (b), we start to see a significant reduction in the number of pairs of concentric rings when applying the baseline filter. This effect becomes more dramatic the larger the number of concentric circles in low variance are considered. Furthermore, as expected, also in the case where the KQ85 mask is applied (red lines) we find less multiple concentric circles than in the case of the GPO mask.
In the case of 5 concentric circles, as shown in Fig.~\ref{fig:n-circlesWMAP} (e), one basically finds no such situations after applying the baseline filter, but there are a few left in the case of the KQ85 mask and over 200 without the removal of any of the spots of unusual brightness! This brings the point home that sets of concentric circles of low variance, as identified by the GP method, should not be seen as evidence for the CCC model. 

%In the following section we will perform numerical CMB simulations and compare the statistical results on concentric rings of low variance following the GP method, with the results obtained without a foreground mask, with the results with our baseline mask, and with the results obtained with standard foreground masks. 

\section{CMB simulations}
In this section we address the question of what would constitute a statistically relevant observation of concentric rings of low variance. This is a difficult question because there is only one real CMB, and therefore one has to perform series of simulations of the CMB to obtain statistics. But those simulations are only as good as the assumptions that are used to model the CMB data.

\begin{figure}[h]
    \centering
    \begin{subfigure}[c]{0.49\textwidth}
         \centering
         \includegraphics[width=\textwidth]{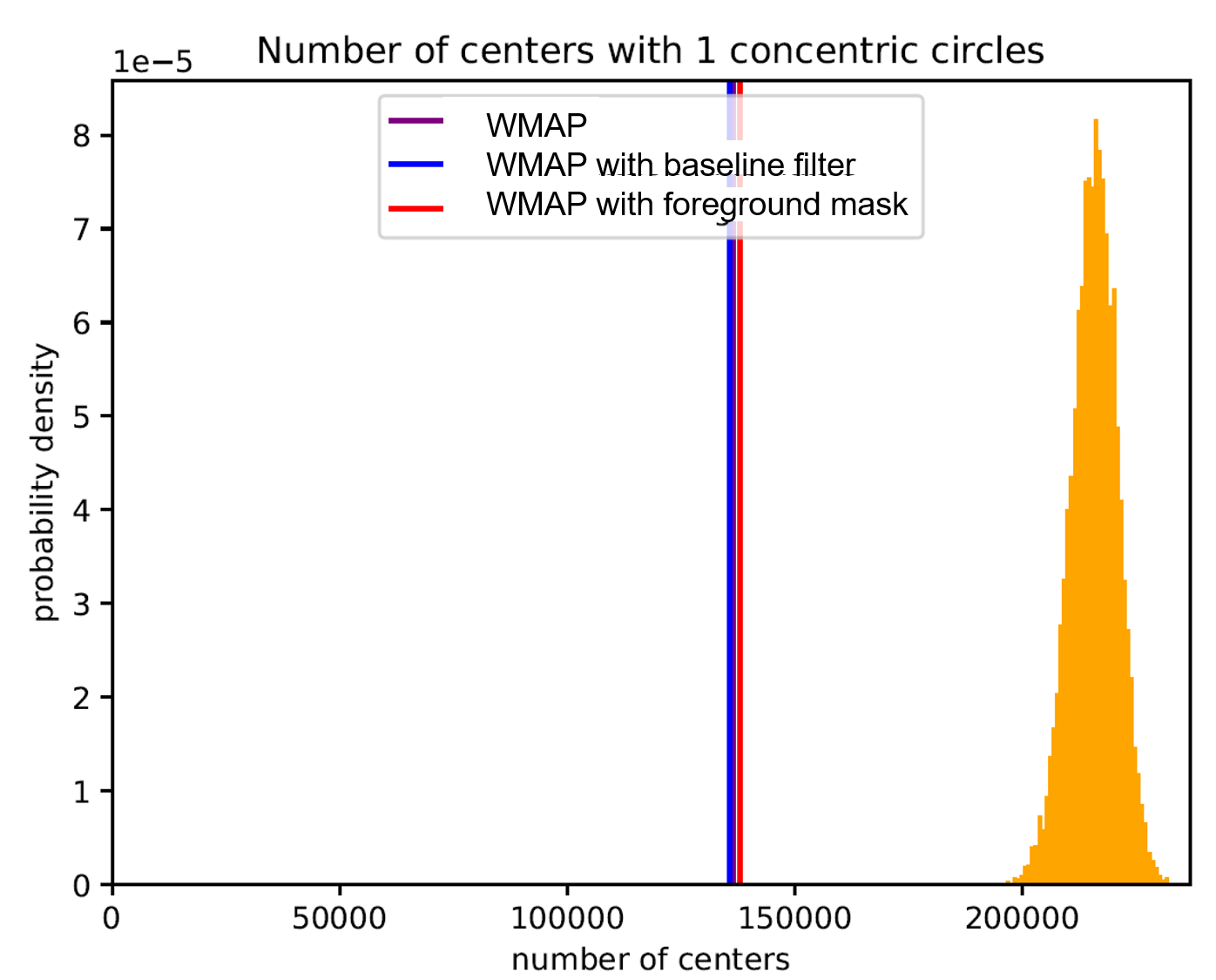}
        %  \caption{$y=5/x$}
        %  \label{fig:five over x}
     \end{subfigure}
        % \caption{Three simple graphs}
        \label{fig:three graphs}
     \begin{subfigure}[c]{0.49\textwidth}
         \centering
         \includegraphics[width=\textwidth]{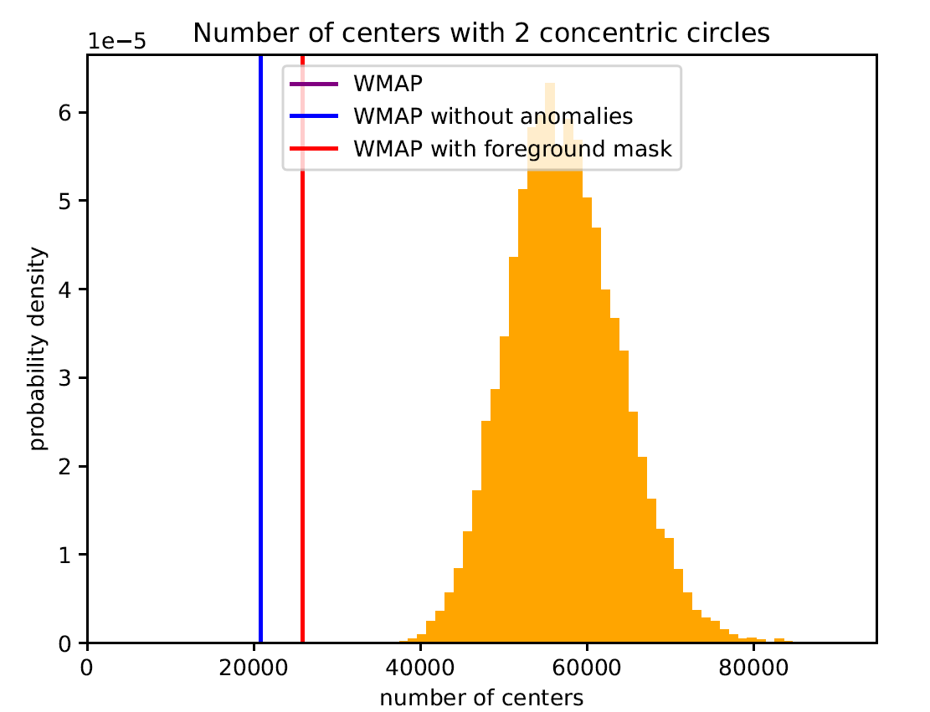}
        %  \caption{$y=5/x$}
        %  \label{fig:five over x}
     \end{subfigure}
        % \caption{Three simple graphs}
        \label{fig:three graphs}
     \begin{subfigure}[d]{0.49\textwidth}
         \centering
         \includegraphics[width=\textwidth]{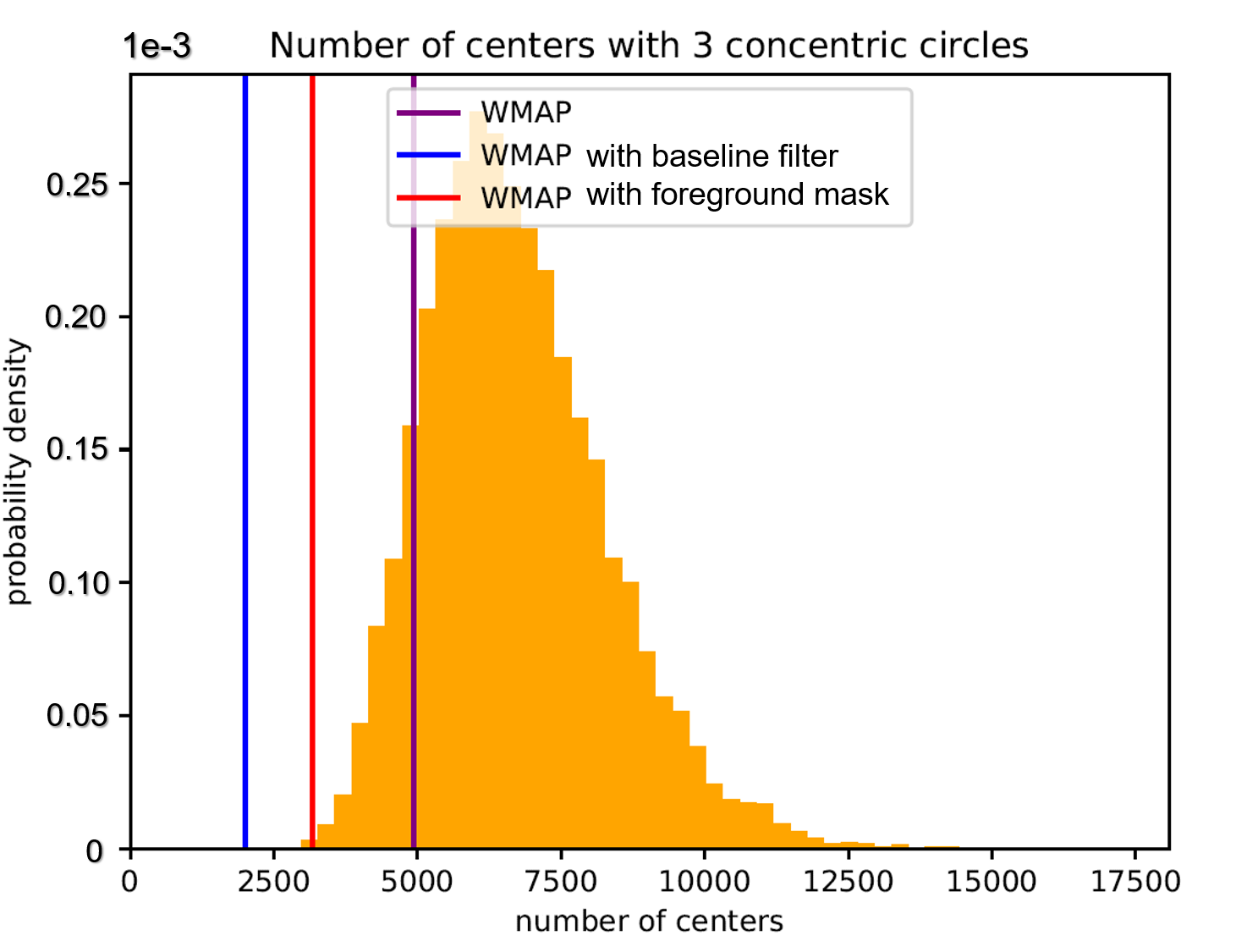}
        %  \caption{$y=5/x$}
        %  \label{fig:five over x}
     \end{subfigure}
        % \caption{2 simple graphs}
        % \label{fig:three graphs}
     \begin{subfigure}[c]{0.49\textwidth}
         \centering
         \includegraphics[width=\textwidth]{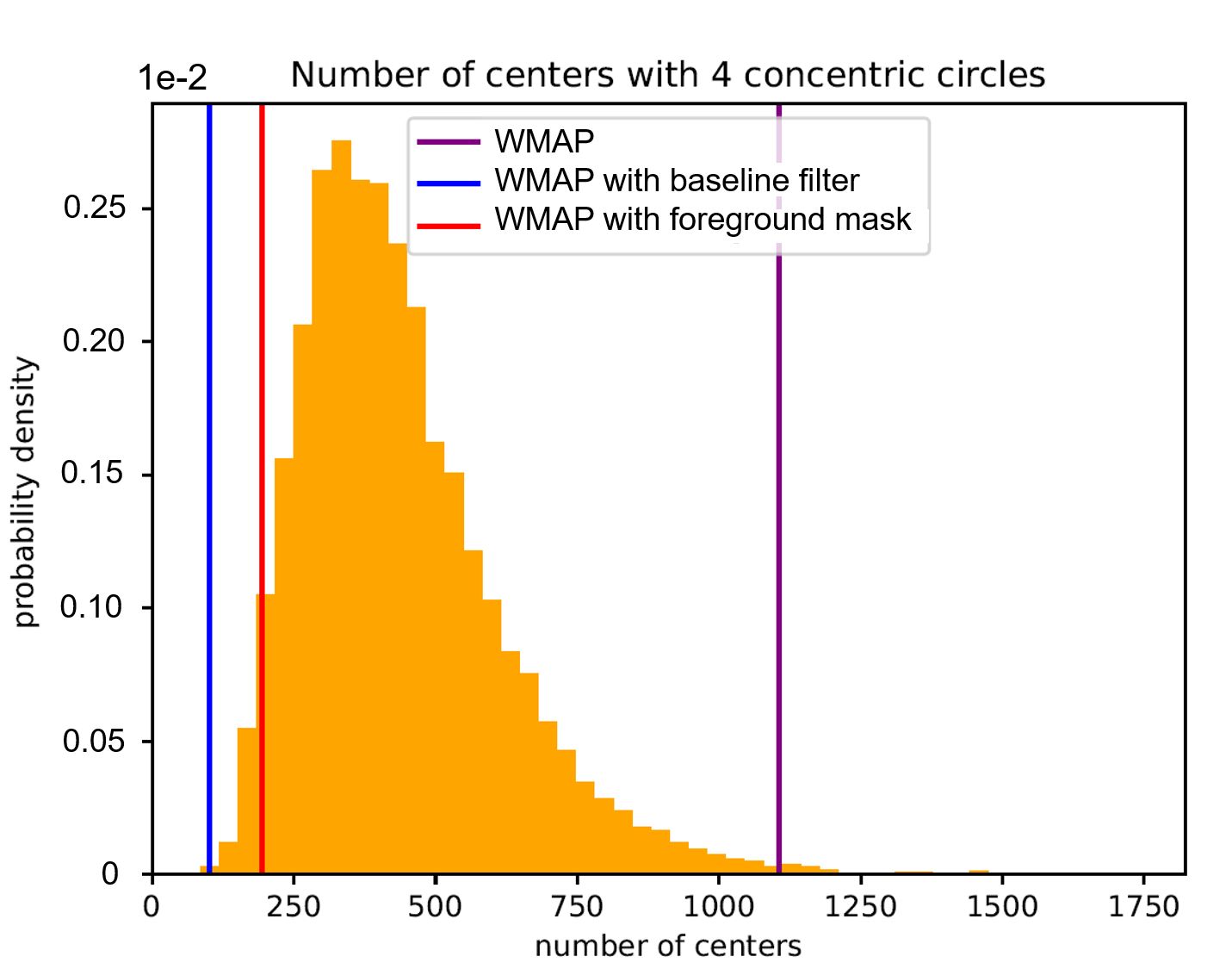}
        %  \caption{$y=5/x$}
         \label{fig:five over x}
     \end{subfigure}
        % \caption{Three simple graphs}
        % \label{fig:three graphs}
     \begin{subfigure}[d]{0.49\textwidth}
         \centering
         \includegraphics[width=\textwidth]{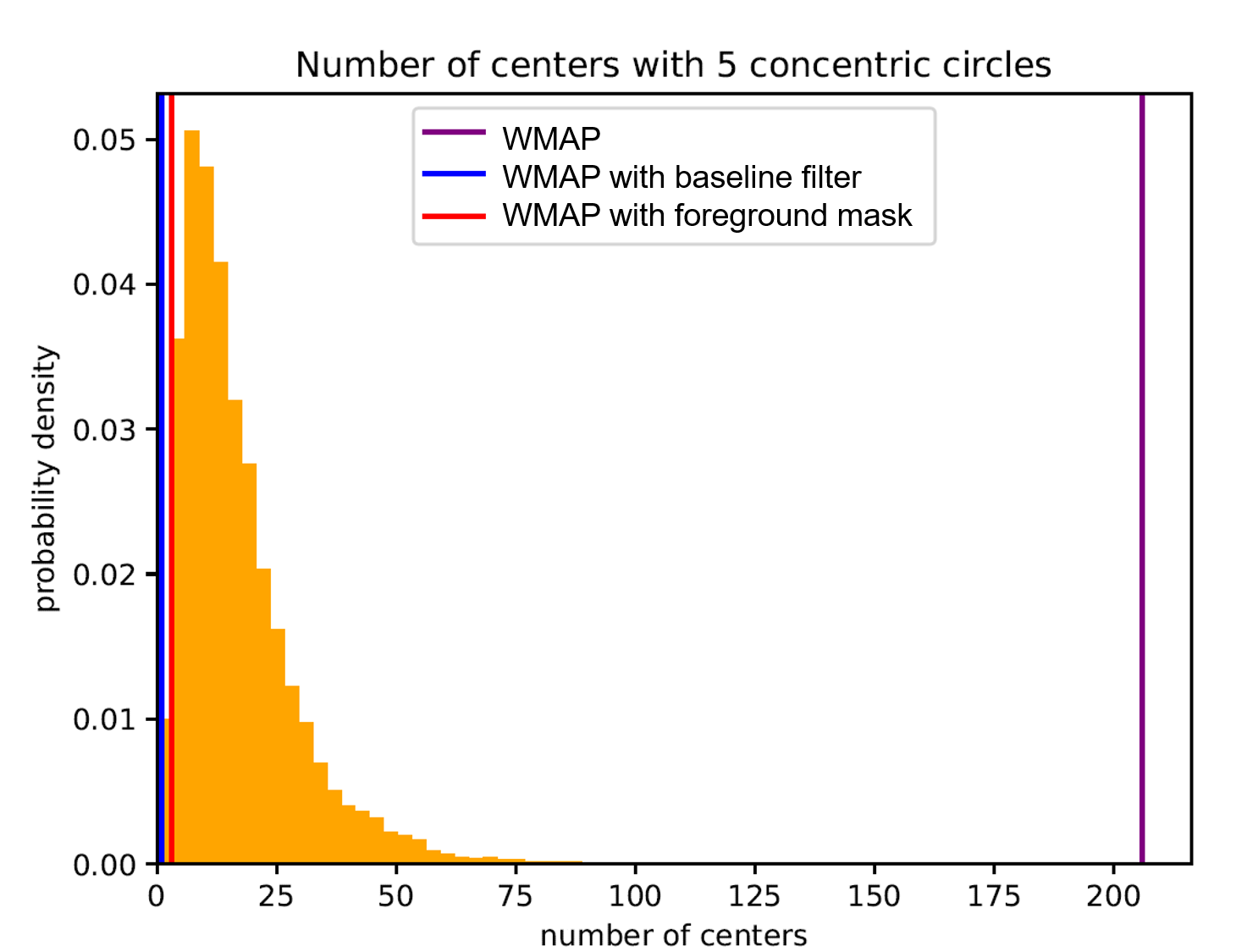}
        %  \caption{$y=5/x$}
         \label{fig:five over x}
     \end{subfigure}
        \caption{The number of concentric circles found with the GP method in WMAP data set with our baseline filter (blue), in the data with GPO mask (purple), and in the data with foreground mask. The yellow histograms correspond to 1000 simulations for every plot, based on Gaussian fluctuations with the CMB power spectrum produced by WMAP "wmap-tt-spectrum-9yr-v5.txt, followed by Galactic plane filtering. }
        \label{fig:n-circlesWMAP}
\end{figure}

The first model that we consider is based on  Gaussian fluctuations with the CMB power spectrum produced by WMAP "wmap-tt-spectrum-9yr-v5.txt", followed by Galactic plane filtering.
The yellow histograms in Fig.~\ref{fig:n-circlesWMAP} correspond to 1000 numerical simulations of the CMB sky.
The simulations produce about 1.6, 2.5, 2.9, 4.0, and 10-100 times the number of single, 2, 3, 4, and 5 concentric circles of low-variance, respectively, using the GP method, than there are present in the real CMB data (blue vertical lines). This indicates that the WMAP CMB with GPO mask is significantly smoother (in regions between 2.5 and 16 degree angular radii) than a model based on Gaussian fluctuations with the CMB power spectrum produced by WMAP "wmap-tt-spectrum-9yr-v5.txt. 

\begin{figure}[h]
    \centering
    \begin{subfigure}[c]{0.49\textwidth}
         \centering
         \includegraphics[width=\textwidth]{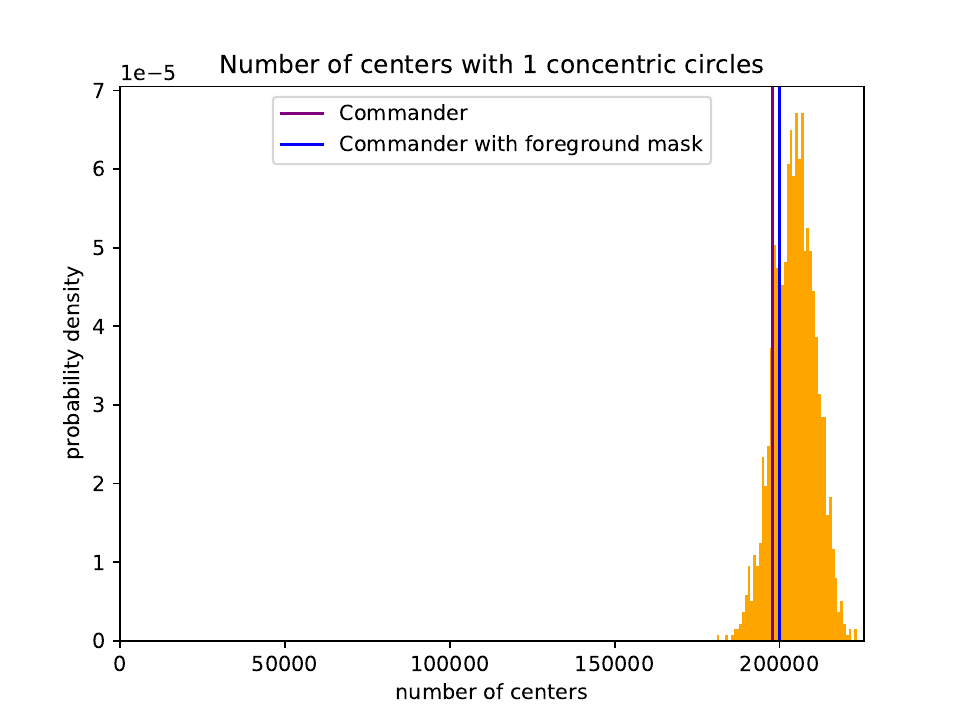}
        %  \caption{$y=5/x$}
        %  \label{fig:five over x}
     \end{subfigure}
        % \caption{Three simple graphs}
        \label{fig:three graphs}
     \begin{subfigure}[c]{0.49\textwidth}
         \centering
         \includegraphics[width=\textwidth]{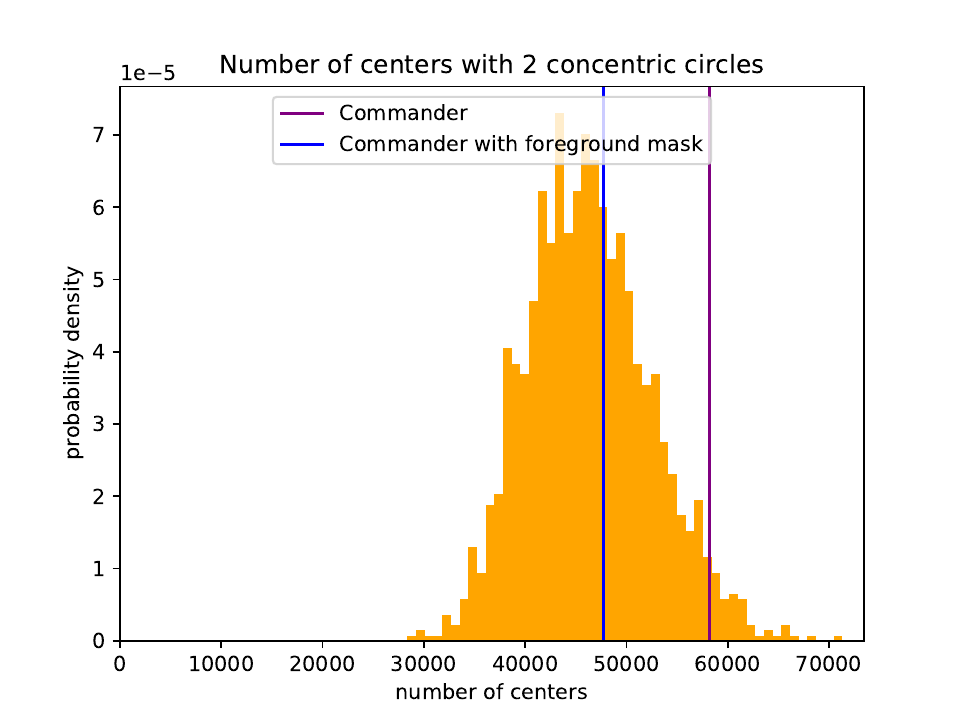}
        %  \caption{$y=5/x$}
        %  \label{fig:five over x}
     \end{subfigure}
        % \caption{Three simple graphs}
        \label{fig:three graphs}
     \begin{subfigure}[d]{0.49\textwidth}
         \centering
         \includegraphics[width=\textwidth]{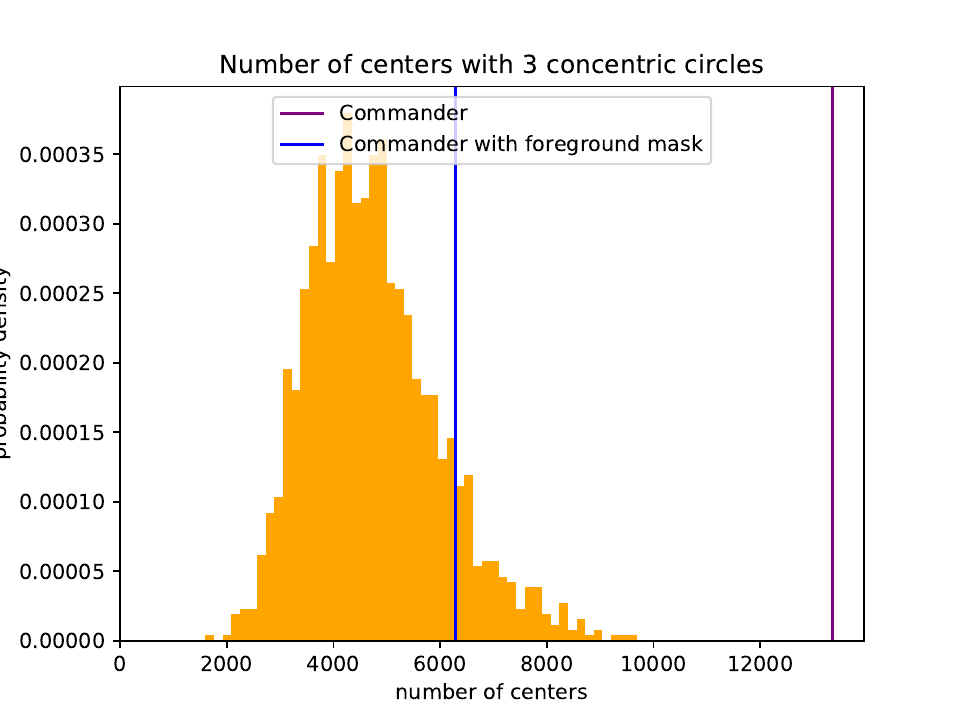}
        %  \caption{$y=5/x$}
        %  \label{fig:five over x}
     \end{subfigure}
        % \caption{2 simple graphs}
        % \label{fig:three graphs}
     \begin{subfigure}[c]{0.49\textwidth}
         \centering
         \includegraphics[width=\textwidth]{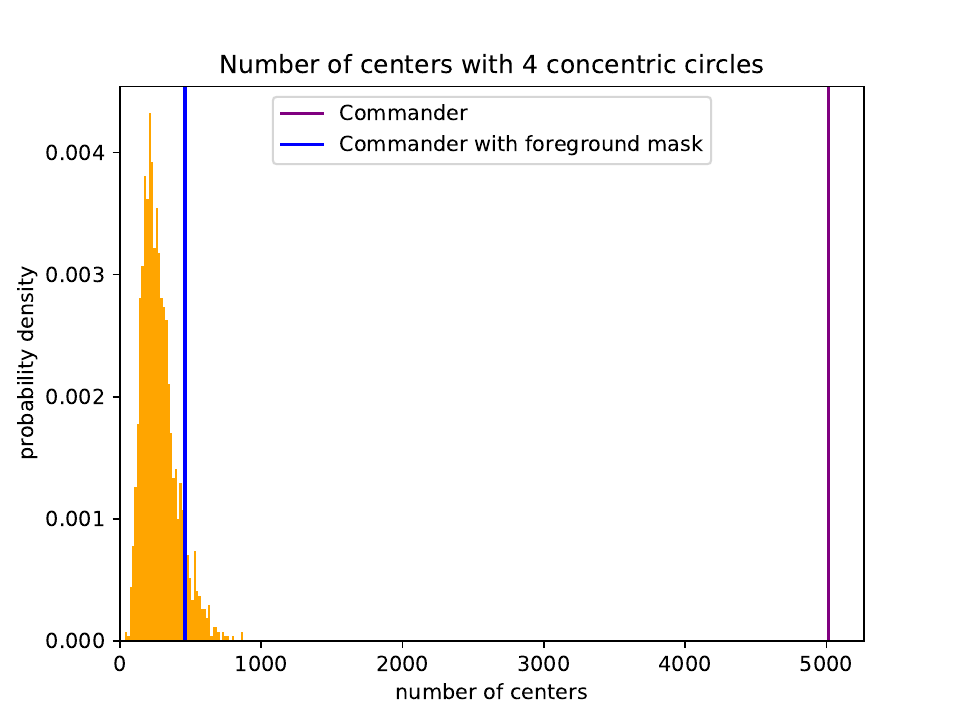}
        %  \caption{$y=5/x$}
         \label{fig:five over x}
     \end{subfigure}
        % \caption{Three simple graphs}
        % \label{fig:three graphs}
     \begin{subfigure}[d]{0.49\textwidth}
         \centering
         \includegraphics[width=\textwidth]{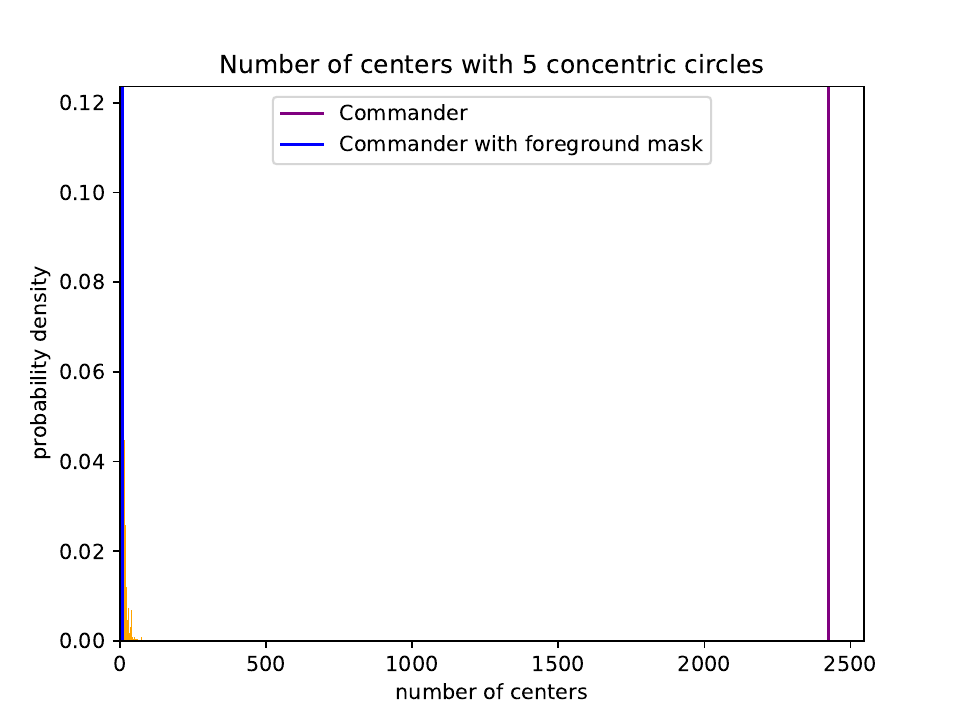}
        %  \caption{$y=5/x$}
         \label{fig:five over x}
     \end{subfigure}
        \caption{The number of concentric circles found in COMMANDER data set with foreground mask "COM-Mask-CMB-common-Mask-Int-2048-R3.00.fits" (blue) and in the data with galactic plane mask only (purple). The orange histograms correspond to 1500 simulations for every plot, based on Gaussian fluctuations with the empirical spectrum "COM-PowerSpect-CMB-TT-full-R3.01.txt" produced by COMMANDER, followed by Galactic plane filtering. }
        \label{fig:n-circlesCOM}
\end{figure}

In order to investigate how robust this observation is against different real CMB data, we repeated the same simulations based on Gaussian fluctuations with the empirical spectrum "COM-PowerSpect-CMB-TT-full-R3.01.txt" produced by COMMANDER, followed by Galactic plane filtering. The orange histograms in Fig.~\ref{fig:n-circlesCOM} show the results for detecting 1 to 5 concentric circles of low-variance in the computer generated CMB data. The result from the real COMMANDER data with Galactic plane filtering only is indicated by the purple vertical bars. 
The real data with the foreground mask "COM-Mask-CMB-common-Mask-Int-2048-R3.00.fits"(thereby removing spots of unusual brightness, including most of the ones we identified in the baseline filter) is indicated by the blue vertical bars. Clearly, in the case of the COMMANDER data the simulations are much better, but far from perfectly, in agreement with the real data (with foreground removal) than in the case of the WMAP data. The main conclusion from the previous section, namely that spots of unusual brightness, combined with the GP method, leads to the erroneous observation of families of concentric rings of low-variance (the purple lines in the plots in Fig.~\ref{fig:n-circlesCOM}), remains very clear (the purple lines are not anywhere close to the simulated data for 3,4, and 5 concentric rings of low-variance).

%The real CMB date, after applying the common foreground mask (thereby removing anomalies, including the ones we identified), indicated by the vertical blue bar in each panel, now lies well within the distributions obtained by the simulations. This means that the best-fit $\Lambda CDM$ and a power spectra Gaussian noise model faithfully reproduces the statistics on observing 1 to 5 concentric rings of low-variance in the CMB, after removal of anomalies.

%and in the data with galaxy plane mask only (purple). The orange histograms correspond to 1500 simulations for every plot, based on Gaussian fluctuations with the emperical spectrum "COM-PowerSpect-CMB-TT-full-R3.01.txt" produced by COMMANDER, followed by Galaxy plane filtering.

\begin{figure}[h]
    \centering 
    \begin{subfigure}[c]{0.49\textwidth}
         \centering
         \includegraphics[width=\textwidth]{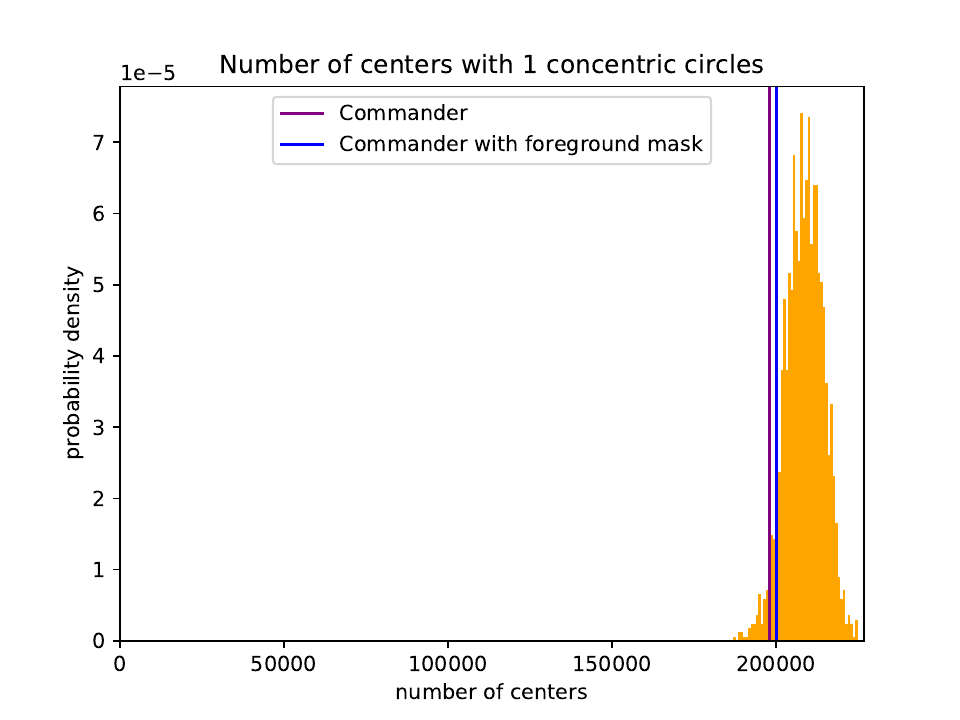}
        %  \caption{$y=5/x$}
        %  \label{fig:five over x}
     \end{subfigure}
        % \caption{Three simple graphs}
        \label{fig:three graphs}
     \begin{subfigure}[c]{0.49\textwidth}
         \centering
         \includegraphics[width=\textwidth]{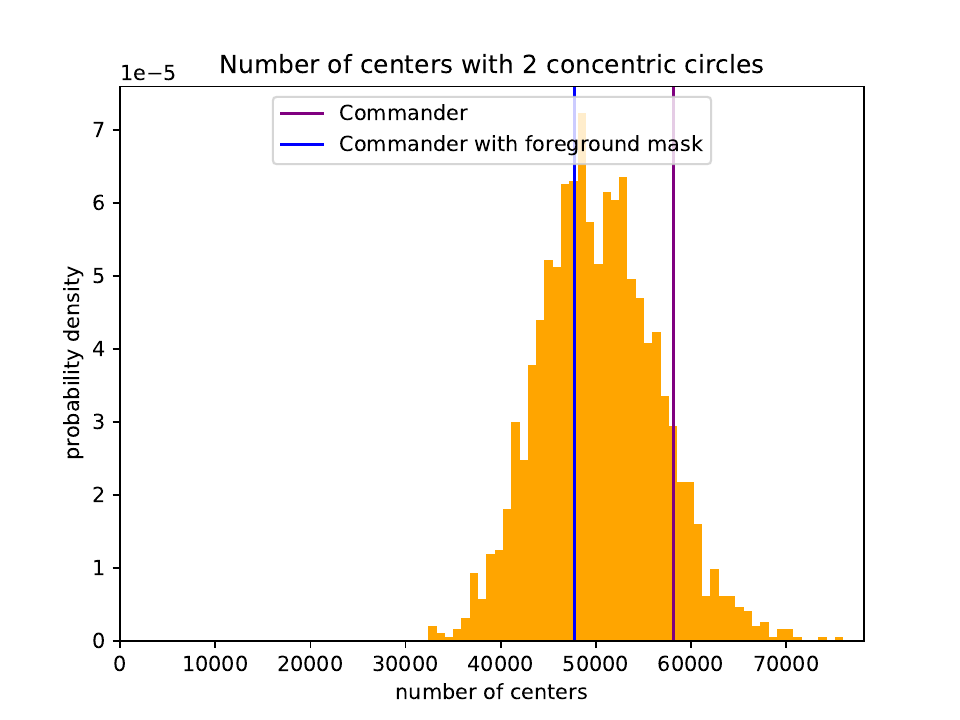}
        %  \caption{$y=5/x$}
        %  \label{fig:five over x}
     \end{subfigure}
        % \caption{Three simple graphs}
        \label{fig:three graphs}
     \begin{subfigure}[d]{0.49\textwidth}
         \centering
         \includegraphics[width=\textwidth]{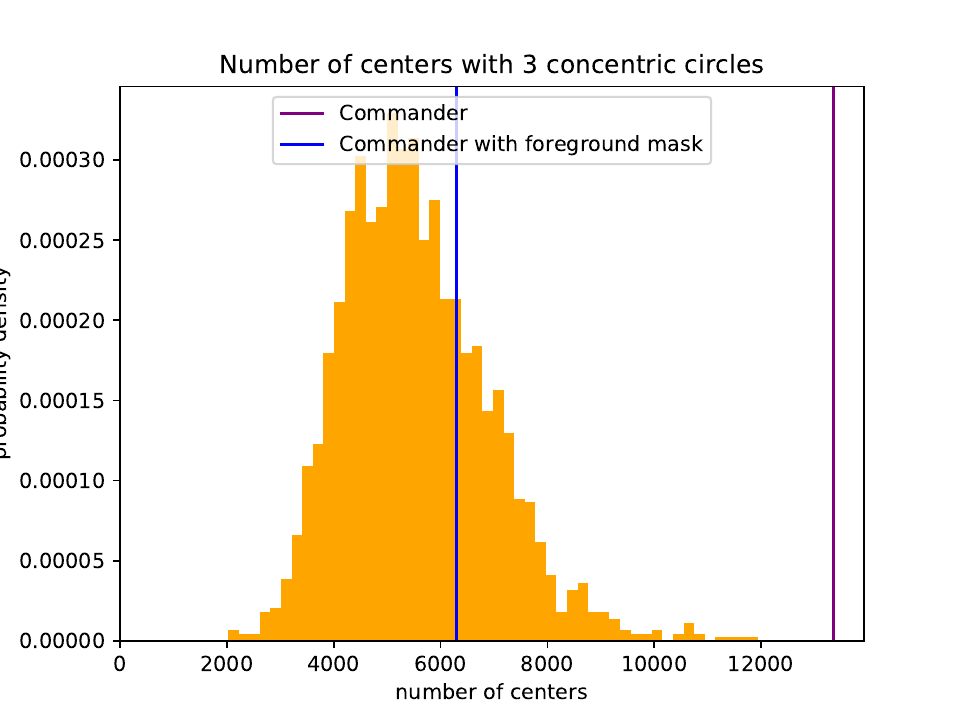}
        %  \caption{$y=5/x$}
        %  \label{fig:five over x}
     \end{subfigure}
        % \caption{2 simple graphs}
        % \label{fig:three graphs}
     \begin{subfigure}[c]{0.49\textwidth}
         \centering
         \includegraphics[width=\textwidth]{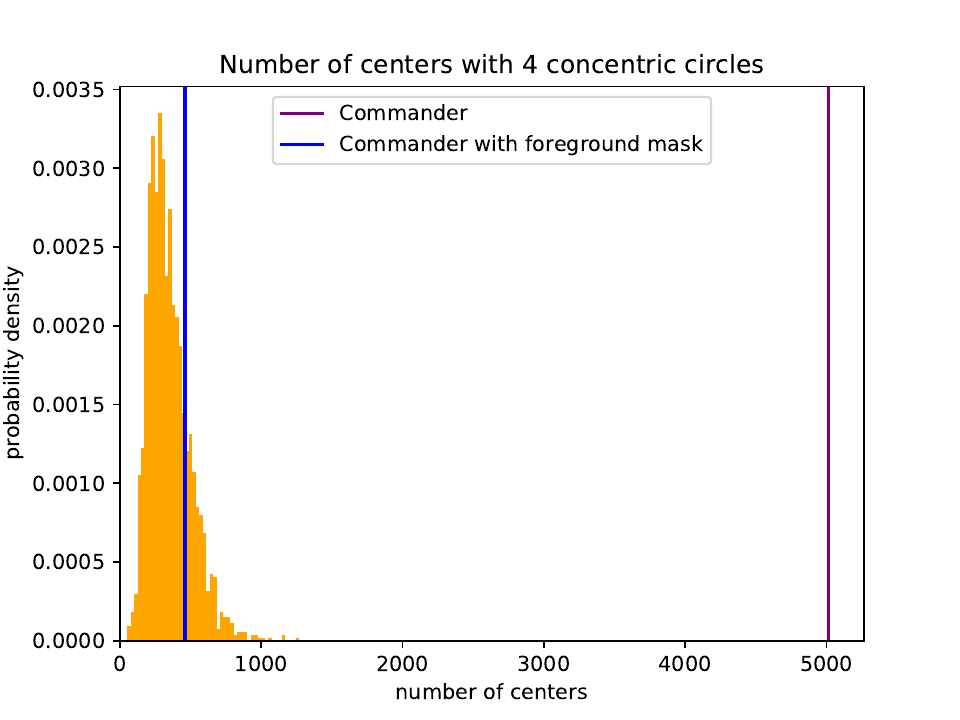}
        %  \caption{$y=5/x$}
         \label{fig:five over x}
     \end{subfigure}
        % \caption{Three simple graphs}
        % \label{fig:three graphs}
     \begin{subfigure}[d]{0.49\textwidth}
         \centering
         \includegraphics[width=\textwidth]{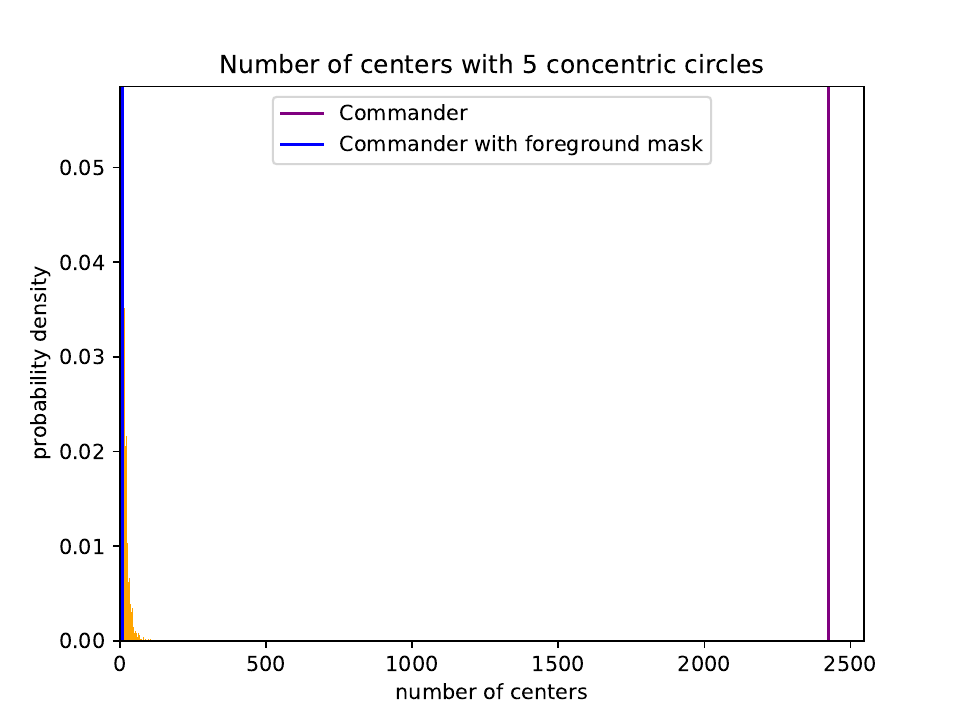}
        %  \caption{$y=5/x$}
         \label{fig:five over x}
     \end{subfigure}
        \caption{The number of concentric circles found in COMMANDER data set with foreground mask "COM-Mask-CMB-common-Mask-Int-2048-R3.00.fits" (blue) and in the data with galactic plane mask only (purple). The orange histograms correspond to 2000 simulations for every plot, based on Gaussian fluctuations with the $\Lambda$CDM spectrum "COM-PowerSpect-CMB-base-plikHM-TTTEEE-lowl-lowE-lensing-minimum-theory-R3.01.txt" produced by COMMANDER, followed by Galactic plane filtering. }
        \label{fig:n-circlesCOMMANDER}
\end{figure}

An important alternative model is that of Gaussian noise with the best-fit $\Lambda CDM$ power spectrum. The orange histograms in Fig.~\ref{fig:n-circlesCOMMANDER} correspond to 2000 simulations, based on Gaussian fluctuations with the $\Lambda$CDM spectrum "COM-PowerSpect-CMB-base-plikHM-TTTEEE-lowl-lowE-lensing-minimum-theory-R3.01.txt" produced by COMMANDER, followed by Galactic plane filtering. The real CMB date, after applying the common foreground mask, indicated by the vertical blue bar in each panel, lies reasonably well within the distributions obtained by the simulations. This means that the best-fit $\Lambda CDM$ and a power spectra Gaussian noise model quite faithfully reproduces the statistics on observing 1 to 5 concentric rings of low-variance in the CMB, after applying the common foreground mask. Again, the main conclusion from the previous section about the role of a few spots of unusual brightness in incorrectly observing abundant sets of concentric circles of low-variance using the GP method is expressed in Fig.~\ref{fig:n-circlesCOMMANDER} by the purple vertical lines. 

\section{\label{sec:level2}A Machine Learning Search of Hawking Points}
We now turn to the investigation of possible Hawking points in the CMB, as predicted by CCC. This is a type of problem, searching for a well-defined feature in a large and noisy data set, that is ideally suited for a machine learning approach. We build upon experience in machine learning approaches to reinforcement learning \cite{vlad}.
Using a GPU 40 TFlops supercomputer and our adopted machine learning software HawkingNet (see Appendix \ref{sec:Appendix A}), we first trained a neural network in distinguishing between the presence and absence of artificially introduced Hawking points at the center of small patches in the sky, and subsequently performed this task on 50 million such centers in the real CMB sky. 

We implement two kinds of training simulations for the CMB sky: one based on a Gaussian noise model %(Figure ~\ref{fig:HP_200}) 
and another on a Pseudo-synthetic model that uses real CMB data before adding artificial Hawking points. %(Figure ~\ref{fig:Pseudo_real_PSN_against_RDS})
The common feature of the two models is that every considered center is enclosed in a $4 \times 4$ degrees patch in the sky - a matrix of the CMB temperature of $128 \times 128$ elements, which approximately corresponds to the Planck data resolutions. For every patch, we subtract the mean temperature in order to avoid temperature bias between all the patches used in our analysis. Here we do have to stay alert to the pitfall of few-pixel spots of unusual brightness in the CMB in a certain patch!
Then the Hawking point simulation signature suggested in Ref.~\cite{Hawking_Penrose} is added and the neural network is trained to optimally identify such points. 
The signature is an added spatially Gaussian-shaped signal with an amplitudes varying from 100 to 400$\mu$K and a width ranging from 0.7 to 1.2 degree in opening angle. After the training, the neural network searched for Hawking points in the real data as well as in CMB simulation data without added Hawking points for comparison.
% The resulted simulations extract around 10,000 patches in favor of the balance between the speed of learning and uniformity of the chosen data by the network, so we can avoid too many patches from the same simulation. 
All the simulations have a resolution Nside=2048, are without smoothing, and have no masks applied other than excluding the galactic plane.

\begin{figure}[hbt!]
    \begin{subfigure}[b]{0.5\textwidth} 
         \includegraphics[width=\textwidth]{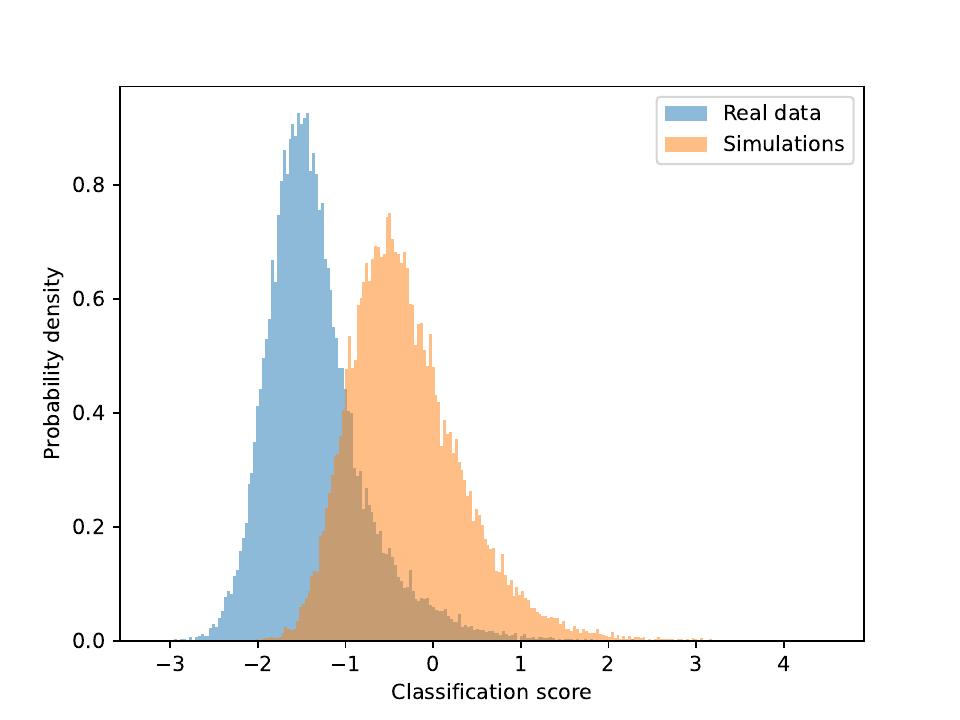}
         %\caption{ The Gaussian simulations with no Hawking points in are generated using HEALPix and the best-fit $\Lambda$CDM power spectrum. }
         %\label{fig:LCDM_Gaussian_ML}
        \caption{}
     \end{subfigure}
     \hfill
     \begin{subfigure}[b]{0.5\textwidth}
         \includegraphics[width=\textwidth]{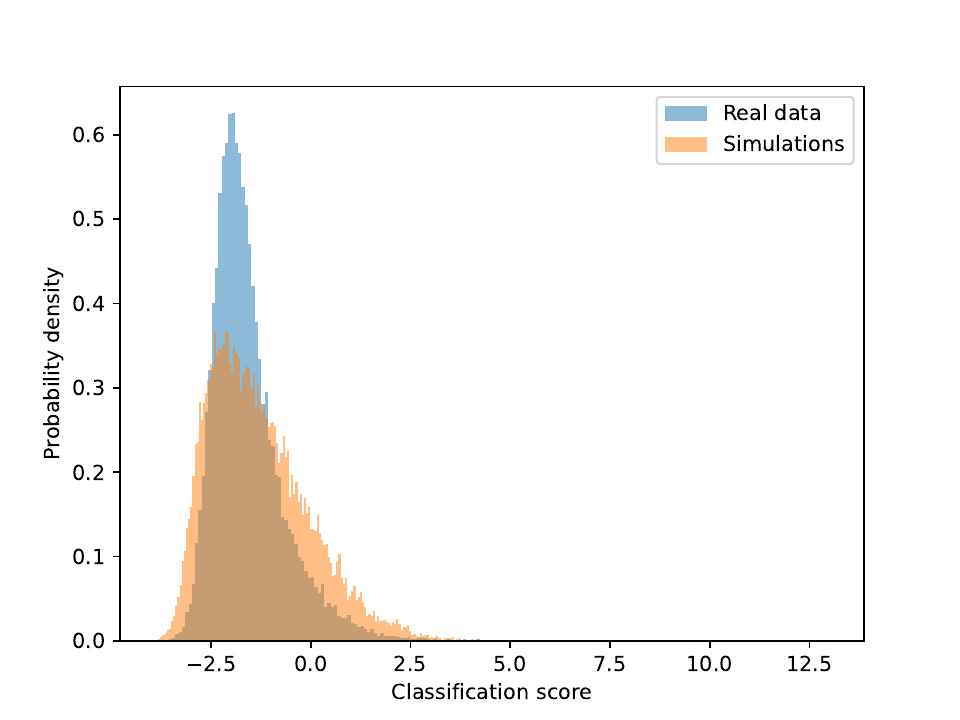}
         %\caption{The Gaussian simulations with no Hawking points generated using HEALPix and "COM PowerSpect CMB-TT-full R301" power spectrum.}
         %\label{fig:COM_Gaussian}
        \caption{}
     \end{subfigure}
       \caption{The Hawking points hypothesis tested on the real data from SMICA Planck and the Gaussian simulations (on a single simulation run that generates 50,000,000 partially overlapping patches in the sky) using ResNet18 ML architecture for 200 $\mu$K added Gaussian amplitude. (a) Blue represents the real data, and orange the Gaussian simulations without added Hawking points, generated using HEALPix and the best-fit $\Lambda$CDM CMB power spectrum from the baseline Planck TT,TE,EE+lowE+lensing. (b) Blue represents the real data, and orange the Gaussian simulations, without additional Hawking points, generated using HEALPix and "COM PowerSpect CMB-TT-full R301" power spectrum. The classification score is a measure of how likely it is that Hawking points are present in patches. 
       }
       %The classification score is a number assigned as the result of the MLE regression by the ML network, depending on how many Hawking points it finds in a scanned patch. Having multiple Hawking points in a patch is exceedingly rare occurrence, so every patch has only one Hawking point.
\label{fig:HP_200}

\end{figure}

%The crucial distinction between the Gaussian and Pseudo-synthetic neural network training methods is \textit{the origin} of the data sets used for training the classifier. 
\textbf{The Gaussian simulation method:} In our Gaussian simulation method, 50 million centers with surrounding patches are assumed to follow a Gaussian global spatial noise distribution and were generated by the HEALPix library which ensures the spatial uniformity for all the points across the sphere \cite{healpix}. HEALPix allows to input the CMB power spectra and generates the (Gaussian) random temperature fluctuations data according to these spectra. We used the best-fit $\Lambda$CDM CMB power spectra from the baseline Planck TT,TE,EE+lowE+lensing and "COM PowerSpect CMB-TT-full R30" - empirically constructed spectra from the real data without any assumption about the CMB model \cite{Planck_Legacy_Archive}. Note that we use Gaussian profiles for both the spatial shape of the Hawking points and for the noise model. After the training of the neural network, both the real CMB and simulated CMB data (without added artificial Hawking points) are subject to a Hawking point search and the results are shown in Fig. \ref{fig:HP_200}

Figure \ref{fig:HP_200}~(a) shows two probability density distributions, the real data from SMICA Planck (blue) and the simulations (orange) using the Gaussian method with the best-fit $\Lambda$CDM CMB power spectrum from the baseline Planck TT,TE,EE+lowE+lensing. Figure ~\ref{fig:HP_200}~(b) shows the density distributions for the real data from SMICA Planck (blue) and the simulations (orange) with the "COM PowerSpect CMB-TT-full R30" power spectrum. The classification score is a number assigned as the result of the MLE regression by the ML network indicating presence of the Hawking points signatures in a scanned patch (see Appendix \ref{sec:Appendix A}); negative/positive values correspond to the likelihood of the absence/presence of Hawking-point-like features. 

What should be expected if this method finds statistically significant Hawking points in a reliable way in the real data? First, one should expect a contribution from statistically insignificant Hawking points ("fake" Hawking points resulting from noise) to overlap with the simulation data (without added Hawking points), and second one should expect the real data to contain contributions at classification scores extending on the positive side of the simulation data. The higher the classification score the higher the confidence of the network in identifying a Hawking point in a specific patch in the sky.

From Fig.~\ref{fig:HP_200} (a) we see that the SMICA Planck CMB data is locally quite different from Gaussian noise with the best-fit $\Lambda$CDM CMB power spectrum from the baseline Planck TT,TE,EE+lowE+lensing. The blue histogram represents the real data and it falls almost completely in the range of negative classification scores, implying that hardly any Hawking-point-like features have been identified, not even statistically insignificant Hawking points (resulting from noise as represented by the orange histogram). This observation suggest that the SMICA Planck CMB data is locally smoother than what follows from the $\Lambda$ CDM model. 
The absence of proper overlap between the simulations and  the bulk of the real data implies that the network was trained on simulation data that was not faithful to the real CMB noise spectrum in the 4 $\times$ 4 degree patches. In such cases one should be suspicious about the validity of the findings of the machine learning search algorithm and better training data is required.

In Fig.~\ref{fig:HP_200} (b), we see a better, but still far from ideal, overlap of the real data and simulation data, based on training of the neural network with the model-free "COM PowerSpect CMB-TT-full R30" power spectrum. Note that the blue histograms in Fig.~\ref{fig:HP_200} (a) and (b) are based on the same real data but give somewhat different histograms because the network was trained with different training data, therefore giving somewhat different classification scores to the same patches in the real data.

\begin{figure}[hbt!]
\centering
\includegraphics[scale=0.6]{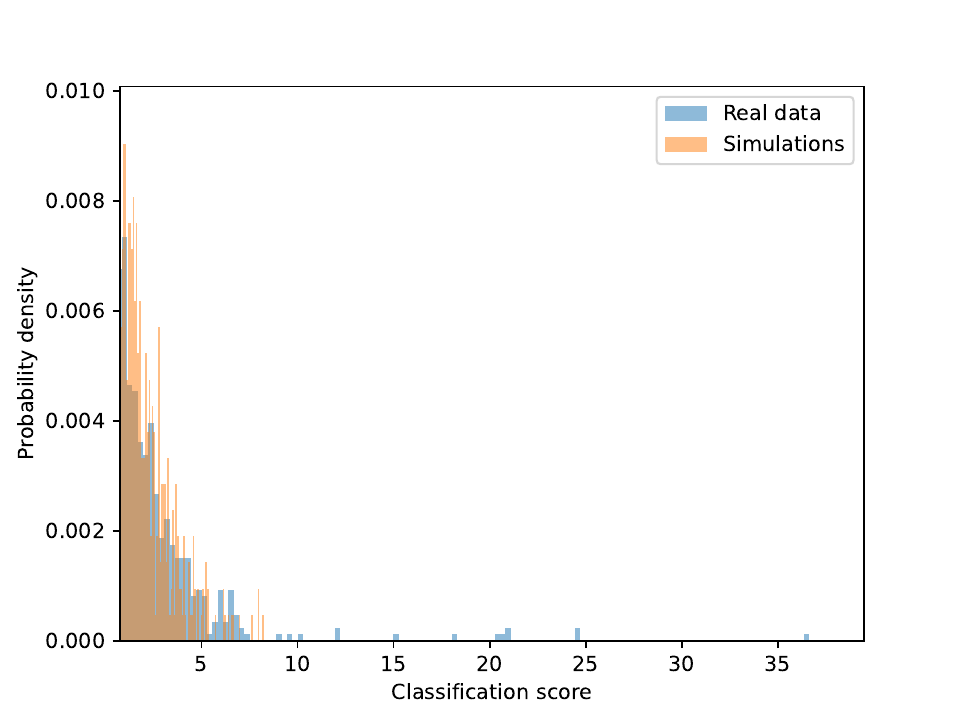}
\caption{The zoomed-in outliers from Fig.~\ref{fig:HP_200} (b) which were classified as Hawking points on SMICA Planck data  with the ML network trained on Gaussian (COM PowerSpect CMB-TT-full R30) simulation data for identifying Hawking points with a 200 $\mu$K amplitude Gaussian signal.}
\label{fig:pp_outliers}
\end{figure}

\begin{figure}[hbt!]
     \centering
     \begin{subfigure}[b]{0.49\textwidth}
         \centering
         \includegraphics[width=\textwidth]{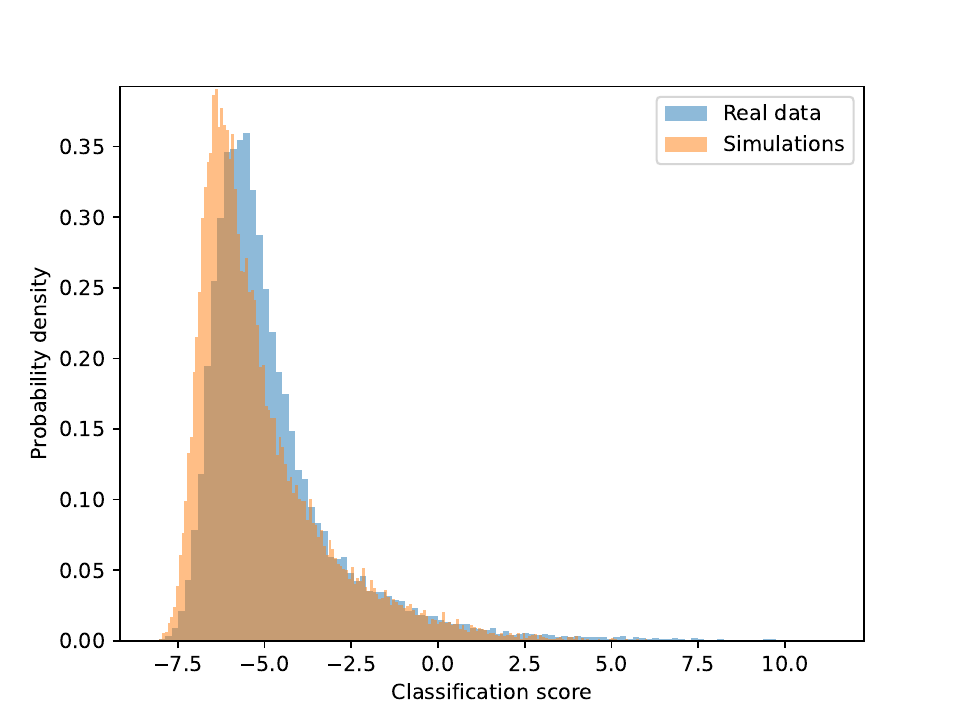}
         \caption{The classification histogram and outlier tail.}
         \label{fig:COMMANDER_full_hist}
     \end{subfigure}
     \hfill
     \begin{subfigure}[b]{0.49\textwidth}
         \centering
         \includegraphics[width=\textwidth]{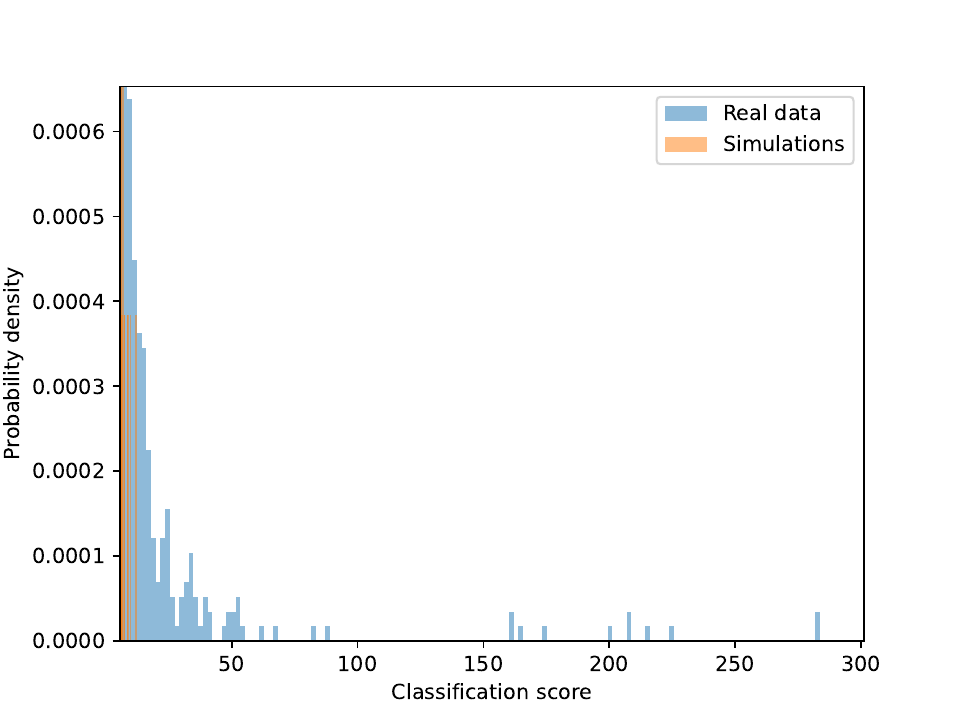}
         \caption{Zoomed in tail of the distribution from (a).}
         \label{fig:COMMANDER_outlier}
     \end{subfigure}

 \caption{The Hawking points hypothesis tested on Planck COMMANDER data using the Gaussian simulation method in the HawkingNet for 200 $\mu$K amplitude. %The classification score is a number assigned as the result of the MLE regression by the ML network indicating presence of the Hawking points signatures in a scanned patch.
 }
        \label{fig:COMMANDER_anomalies}
\end{figure}

Having gained some confidence in this neural network in identifying real Hawking points, which should have classification scores with positive values above the values in the simulations, we zoom in on the region of high classification scores, as shown in  Fig.~\ref{fig:pp_outliers}. We do indeed observe contributions with high classification scores in the real SMICA Planck data. We also used the same neural network to analyse the Planck COMMANDER data as show in Fig.~\ref{fig:COMMANDER_anomalies}. Figure \ref{fig:COMMANDER_anomalies} (a) shows a reasonable overlap between the simulations and real data and Fig.\ref{fig:COMMANDER_anomalies} (b) shows significant contributions, much more pronounced than in the Planck SMICA data, in the real data at high classification scores.

\begin{figure}[hbt!]
\centering
\includegraphics[scale=0.6]{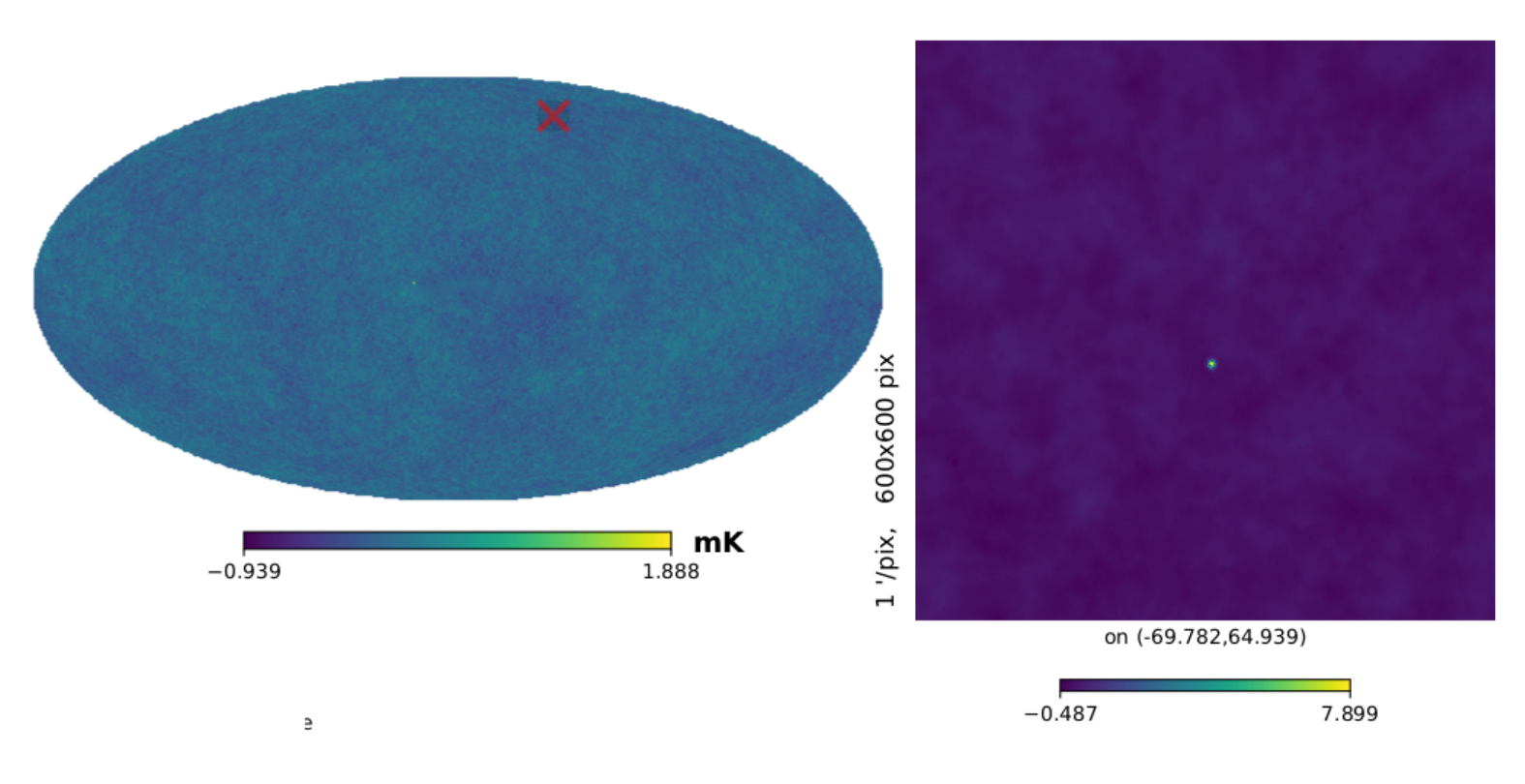}
\caption{The spot of unusual brightness in the SMICA Planck data set that is the reason for all the outliers with high classification scores in Fig.~\ref{fig:pp_outliers}.}
\label{fig:HP_map_outlier}
\end{figure}

Having learned to be extremely cautious in drawing scientific conclusions from searches for features in the CMB, we investigated the specific patches in the sky that corresponded to the high classification scores. In the case of the SMICA Planck data it turned out that there is a single high-temperature spot in the sky, shown in Fig.~\ref{fig:HP_map_outlier}, that misled the neural network in identifying Hawking points in patches of the sky containing this spot. Analyzing the patches in the sky in the Planck COMMANDER data that correspond to the outliers in Fig.\ref{fig:COMMANDER_anomalies} (b) we find 11 spots, one of which is shown in Fig.~\ref{fig:fig27} and the remaining 10 listed in Appendix \ref{sec:Appendix B}, Figs.~\ref{fig:figureApendix} (a)--(j)  with temperature scales with respect to the overall CMB mean temperature of 2.725 K.

We observe that all these spots are single to few-pixel low temperatures spots (between 2.6763 K (Fig.~\ref{fig:fig27}) and 2.7203 K). Given that we haven't observed such cold spots in the SMICA data set, they could be artifacts of the original data processing for the COMMANDER data, or they are real and have been removed from the SMICA data by data processing. 

The high temperature spot  in the Planck SMICA data was apparently picked up by the neural network to have a large likelihood of representing a Hawking point (with the bright spot near the center of the patches). In the case of the Planck COMMANDER data the low temperature spots must have reduced the average background temperature of a patch and thereby favoring those patches (with a cold spot near the boundary of the patches) to detect a raised temperature at the center, suggesting the presence of a Hawking point. 

We conclude that, just as in the case of the search for rings of low variance, spots of unusual brightness deceive the search algorithms in identifying Hawking points, and no statistically significant Hawking points have been identified. 

% In generally, one may increase the parameters of the HawkingNet to tune the neural network to explore the coolest points or the hottest features in the CMB sky, which would be possibly useful for searching the anomalies in the large data sets for the astrophysics community.

\begin{figure}[hbt!]
\centering
\includegraphics[scale=0.6]{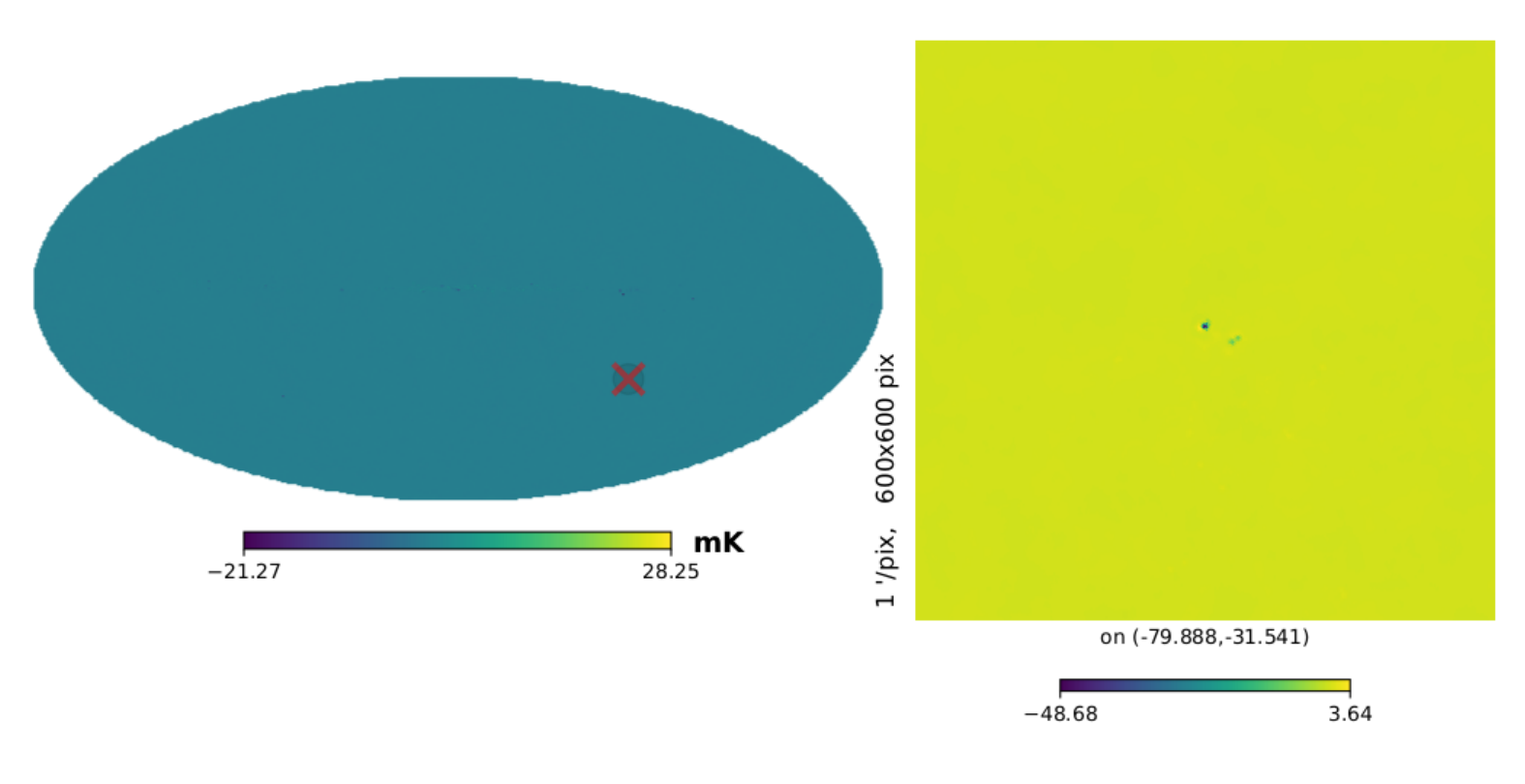}
\caption{One of the 11 spots of unusual temperature found by the HawkingNet in the Planck COMMANDER data. }
\label{fig:fig27}
\end{figure}

Can we improve our machine learning approach for finding Hawking points? As pointed out above, the neural network training starting from Gaussian noise with the best-fit $\Lambda$CDM CMB power spectrum from the baseline Planck TT,TE,EE+lowE+lensing was not very successful, and the training starting from the "COM PowerSpect CMB-TT-full R30" power spectrum was better, but not optimal. Given the fact that the later approach did identify patches with potential Hawking points, which turned out to be indeed regions that likely will be identified as potential Hawking points by any other searching algorithms, unless the single or few pixel spots of unusual brightness are filtered out from the data, indicates that the ML algorithm served it's purpose fairly well. Therefore we can at this point already make the claim that with the current CMB data resolution, Hawking points with a magnitude in the range from 100 to 400 $\mu$K amplitude and spatial Gaussian extend between 0.7 and 1.2 degrees, and small distortions thereof, can be ruled out. We did however explore one more neural network training method, the pseudo-synthetic approach, which could support the above claim with higher confidence.      

\textbf{The pseudo-synthetic approach:} In this method the classifier was trained on real CMB SMICA Planck data with and without added Hawking points, hence the name pseudo-synthetic. The choice of the Planck SMICA set over COMMANDER was motivated by the smaller amount of spots of unusual brightness in the data. We considered 500,000 patches of $4 \times 4$ degree size from both the real data Northern and Southern hemispheres. 

A concern that one could raise for this approach is that the patches of the real data might already contain real Hawking points and if we train the network on such patches with and without additional artificial Hawking points this might lead to a corrupt classification score. This scenario has been studied in the machine learning literature with the conclusion that, as long as the set of training data  (before adding or not adding the feature of interest to individual element of the training data) do not already contain the feature of interest in more than 50\% of the cases of the individual elements of the training data, the training will still converge to a reliable classifier.
For our case this means that we work under the assumption that in the 500,000 patches of the Northern (or Southern) Hemisphere use in the training procedure less than half of them contains a real Hawking point. This seems indeed a very reasonable assumption. 

\begin{figure}[hbt!]
     \centering
     \begin{subfigure}[b]{0.45\textwidth}
         \centering
         \includegraphics[width=\textwidth]{pics/Figure_north_train_south_test.pdf}
         \caption{}
         \label{}
     \end{subfigure}
     \hfill
     \begin{subfigure}[b]{0.45\textwidth}
         \centering
         \includegraphics[width=\textwidth]{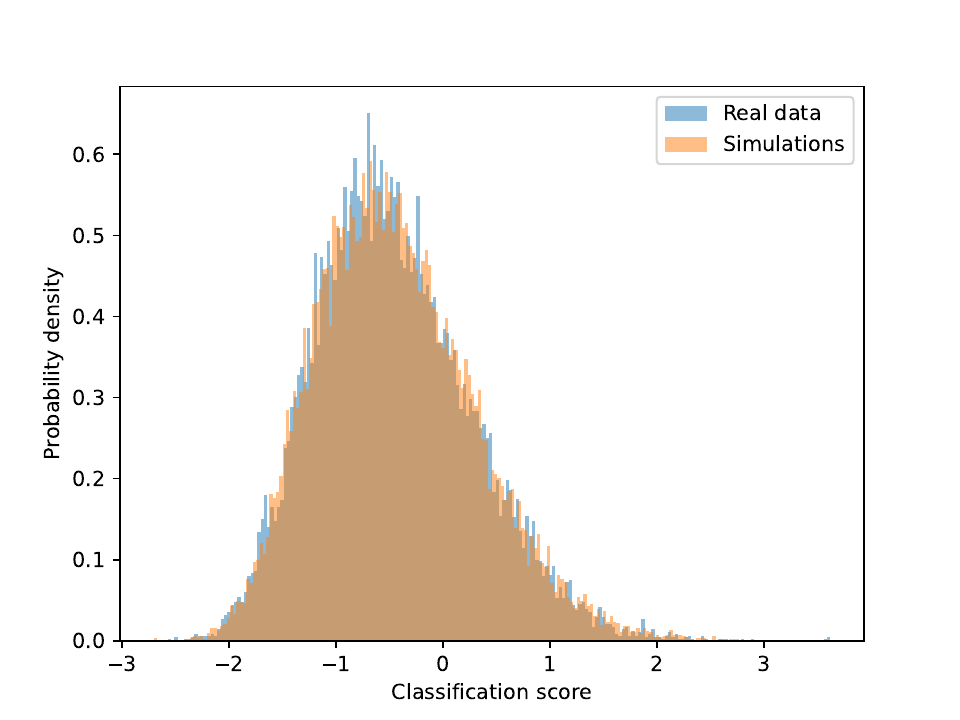}
         \caption{}
         \label{}
     \end{subfigure}
\caption{The Hawking points statistical test on the \textbf{Pseudo-synthetic} SMICA data trained on the (a) northern CMB hemisphere compared to the real data from the southern hemisphere (blue distribution), (b) southern CMB hemisphere compared to the real data from the northern hemisphere (blue distribution), and compared to the simulations based on "COM PowerSpect CMB-TT-full R301" power spectrum (orange distribution).}
\label{fig:Pseudo_real_PSN_against_RDS}
\end{figure}

\begin{figure}[hbt!]
     \centering
     \begin{subfigure}[b]{0.45\textwidth}
         \centering
         \includegraphics[width=\textwidth]{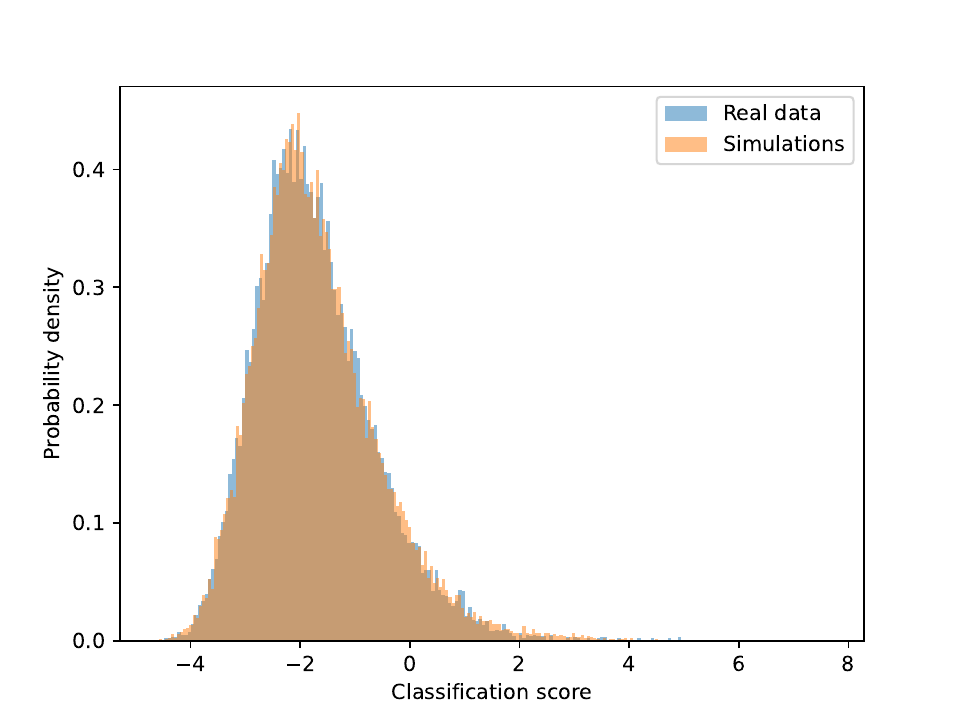}
         \caption{}
         \label{}
     \end{subfigure}
     \hfill
     \begin{subfigure}[b]{0.45\textwidth}
         \centering
         \includegraphics[width=\textwidth]{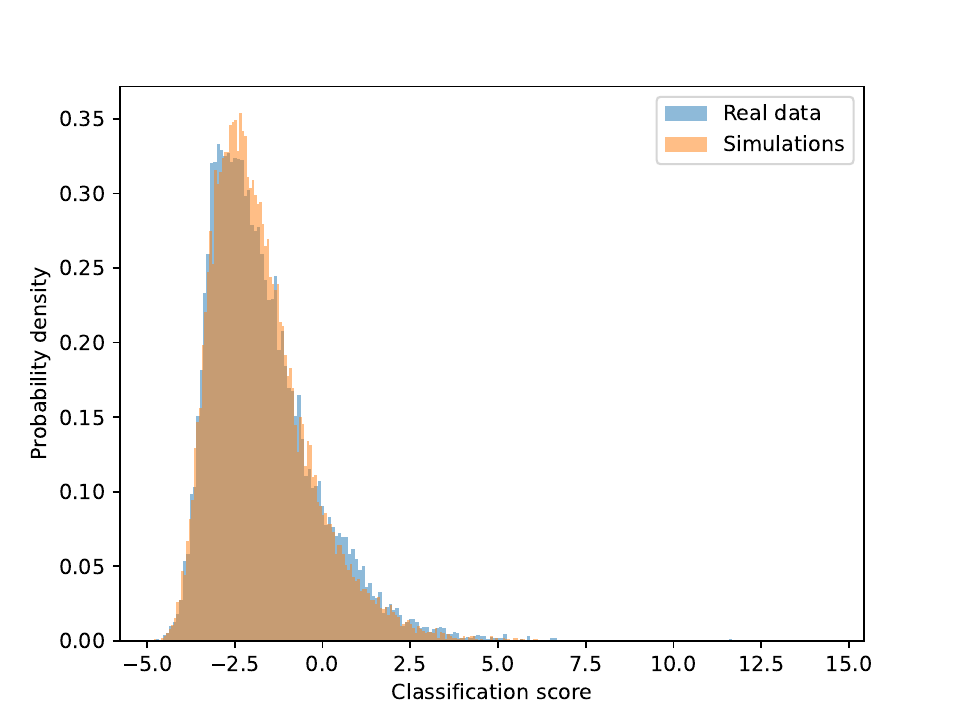}
         \caption{}
         \label{}
     \end{subfigure}
\caption{The Hawking points statistical test on the \textbf{Pseudo-synthetic} COMMANDER data trained on the (a) northern CMB hemisphere compared to the real data from the southern hemisphere (blue distribution), (b) southern CMB hemisphere compared to the real data from the northern hemisphere (blue distribution), and compared to the simulations based on "COM PowerSpect CMB-TT-full R301" power spectrum (orange distribution).}
\label{fig:Pseudo_real_PSN_against_RDS-COMMANDER}
\end{figure}

After training of the classifier on the Northern (Southern) Hemisphere, we applied the classifier to the Southern (Northern) hemisphere date and to the same Gaussian simulation data without Hawking points as used in Fig. ~\ref{fig:HP_200} (b). The resulting classification scores on the real data is shown in blue in Fig.~\ref{fig:Pseudo_real_PSN_against_RDS} for SMICA and  Fig.~\ref{fig:Pseudo_real_PSN_against_RDS-COMMANDER} for COMMANDER, and the classifier score on the simulation data is shown in orange.

\begin{figure}[hbt!]
\centering
\includegraphics[scale=0.8]{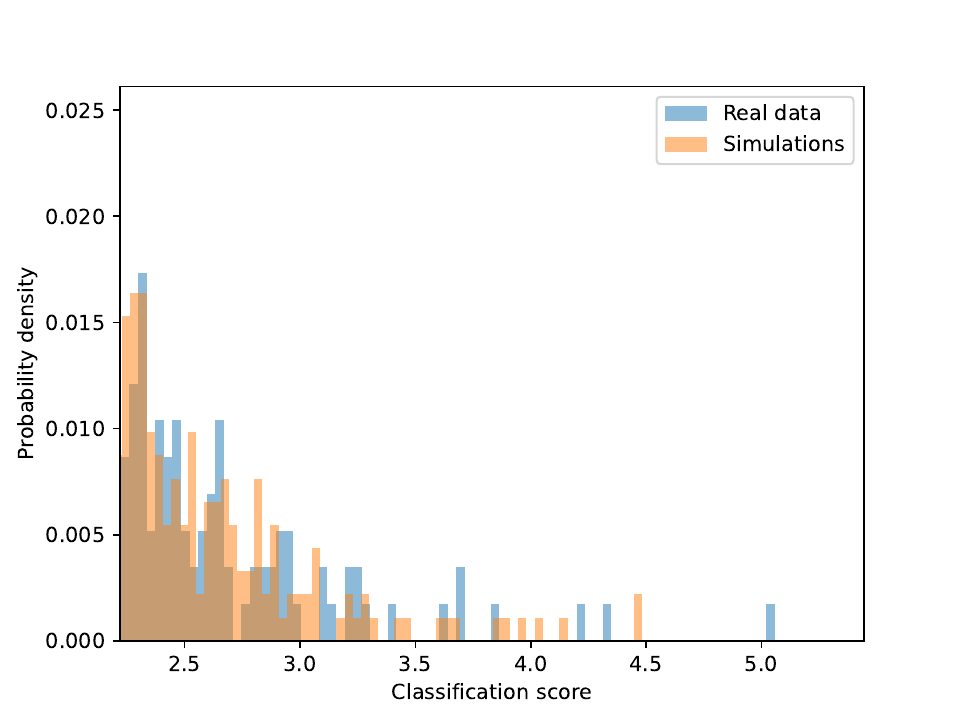}
\label{fig:smica_train_south_test_north_zoom}
\caption{The Zoomed-in  tail of the Hawking points statistical test on SMICA for the \textbf{Pseudo-synthetic} data trained on the southern CMB hemisphere compared to the real data from the northern hemisphere with few-pixel spots of unusual brightness removed.}
\label{fig:smica_train_south_test_north_zoom_SMICA}
\end{figure}

\begin{figure}[hbt!]
\centering
\includegraphics[scale=0.6]{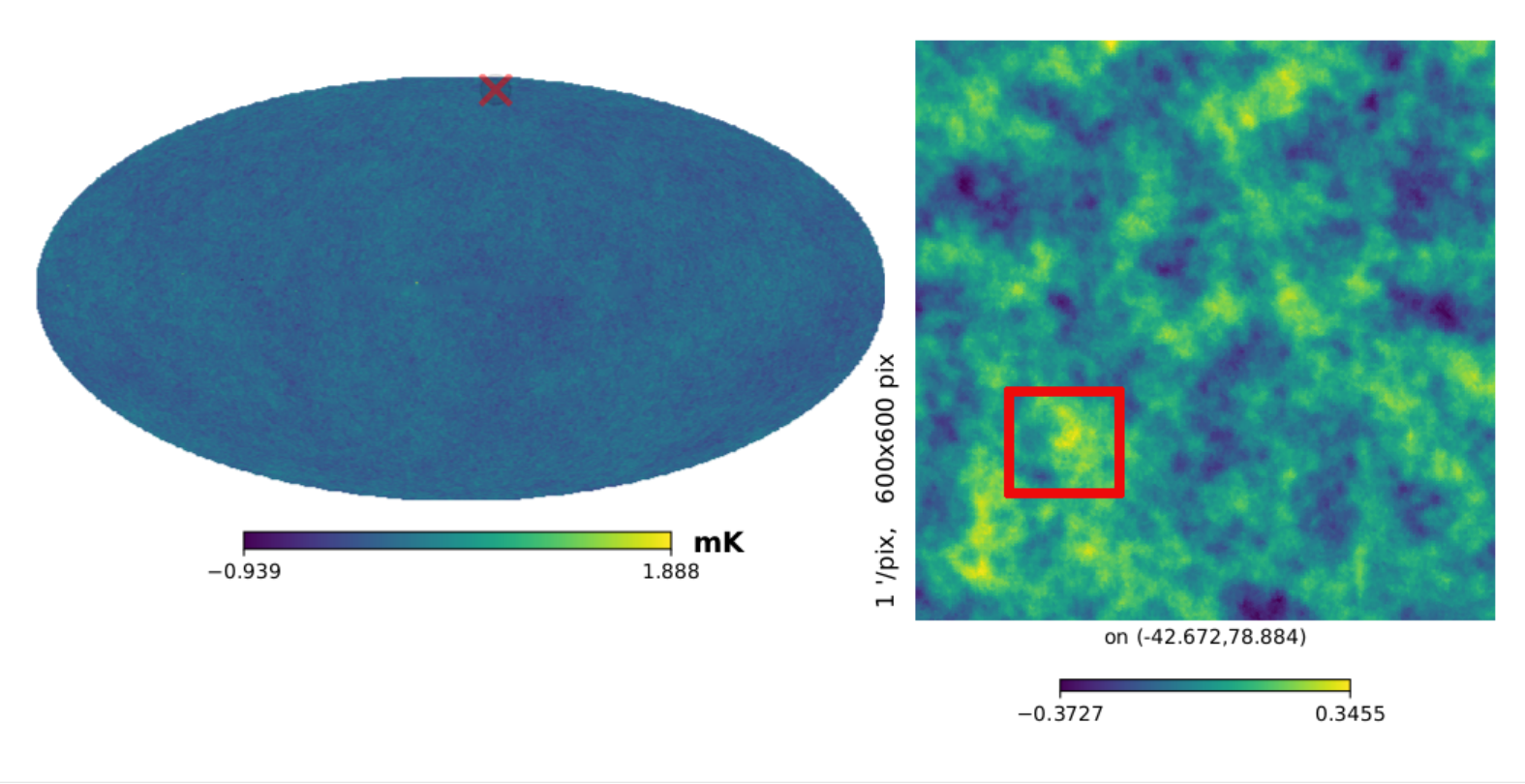}
\caption{The location of the found outlier in Fig. \ref{fig:smica_train_south_test_north_zoom_SMICA} }
\label{fig:smica_outlier_location}
\end{figure}

Next we repeated the statistical test on pseudo-synthetic data set similarly to Fig.~\ref{fig:Pseudo_real_PSN_against_RDS}, but excluded the very few spots of unusual brightness/temperature in the Planck SMICA data and plotted the zoomed in classification score in the positive tail of the distribution, shown in Fig.~\ref{fig:smica_train_south_test_north_zoom_SMICA}. The real data and the simulation data now almost perfectly overlap, confirming and strengthening our conclusion that Hawking points with a magnitude in the range from 100 to 400 $\mu$K amplitude and spatial Gaussian extend between 0.7 and 1.2 degrees, and small distortions thereof, cannot be found in the currently available CMB data. To be thorough, we also investigated the real date patch corresponding to the highest classification score of approximately 5 (See Fig.~\ref{fig:smica_train_south_test_north_zoom_SMICA}). This patch is shown in Fig.~\ref{fig:smica_outlier_location} and is according to our machine learning search the spot in the sky that comes closes to a Hawking point. It certainly does not have the clear Hawking point features and is indeed well explained by random fluctuations.

%\begin{figure}[hbt!]
%\centering
%\includegraphics[scale=0.8]{pics/patch.pdf}
%\caption{The zoomed-in red-boarded patch of Fig. %\ref{fig:smica_outlier_location} }
%\label{fig:smica_outlier_location_zoomed_patch}
%\end{figure}

% \subsection{Gaussian and $\Lambda$CDM Models for the CMB}
% Why CMB is not Gaussian (all the figures) and why it doesn't follow $\Lambda$CDM fit (Fig ~\ref{fig:LCDM_Gaussian_ML})

\textbf{Searching for Concentric Circles Around spots of unusual brightness:}
The CCC predictions of low variance circles are very qualitative because details would depend on the very complex dynamics of interfering and dissipating gravitational shock waves in the plasma at the moment of decoupling of light (now the CMB) from matter, 378,000 years after the beginning of our universe. Also the Hawking point predictions are very qualitative. This led us to a final attempt to find a signature for the CCC hypothesis, namely by consider the identified spots of unusual brightness as possible Hawking points and search for concentric circles of slightly lower temperatures, or circles of low-variance, around them \cite{CCC_Penrose_Gurzadyan}.

We repeated the low-variance circles analysis around the identified spots with small steps of 0.2 degrees, and did not find statistically significant circles. %around these anomalies with the currently available  CMB data resolution. %Higher data resolution would provide larger statistics for the Hawking anomalies classified by the neural network and hence more accurate statistical tests. %So far, having less than 20 anomalies for the data set with over million points for both SMICA and COMMANDER data sets, is not enough to test the Hawking signal hypothesis. We hope to repeat this analysis on more powerful supercomputers and higher resolutions data sets available in future.

\section{Discussion and Conclusion}

We presented a high-resolution search for low-variance temperature circles  in the Planck and WMAP Cosmic Microwave Background (CMB) data, and introduced HawkingNet, our machine learning open-source software based on a ResNet18 algorithm, to search for Hawking points in the data. We found that few-pixel sized spots of unusual brightness/temperature erroneously lead to regions with  near-concentric low-variance circles as well as Hawking points when applying the search criteria used in previous works. Given the fact that our analysis (based on the use of a supercomputer and a powerful machine learning approach), unlike previous studies, uses the maximum available resolution of CMB data, we conclude that no statistically relevant sets of concentric low-variance circles and Hawking points can be found in the currently available CMB data. 

The absence of such statistically-significant distinct features in the currently available CMB data does, of course, not disprove the CCC model but implies that higher resolution CMB data and/or refined CCC based predictions are needed to pursue the search for CCC signatures. 
%Perhaps highly deformed circles and distorted Hawking points can still be extracted but only if there are clear theoretical predictions for this. 
% Perhaps the density of gravitational shock waves from merging black holes in the previous eon has been so high that a complex gravitational interference pattern (similar to optical speckle patterns) arose instead of circular patterns 
%we should expect to observe the result of many interfering spherical waves, leading to a complex interference pattern. Such a complex interference pattern at the moment of decoupling  would require a very different characterization method than a search for clearly distinguishable ring patterns and Hawking points in the CMB.
%Our work shows that  previous attempts on identifying signatures of conformal cyclic cosmology are inconclusive, and suffer from the dramatic impact of CMB anomalies on the data analyses. Unlike our study, previous studies were significant constrained by limited computational power, and as such had largely overlooked the decisive role of anomalies at the level of single pixels in the CMB date.
Our analyses can be expended upon in order to test predictions from various cosmological models. We provide the open-source code for our machine learning method and hope that this will inspire more research on testing cosmological models: \url{https://github.com/winger/hawking_net}. 
%Higher resolution CMB data and/or new predictions from CCC seem to be needed to continue the search for signatures of the CCC model. 

\acknowledgments
We would like to thank Jan Willem Dalhuisen for fruitful discussions and  express our admiration for Sir Roger Penrose for proposing and explaining the intriguing CCC model.

\appendix
\section{The HawkingNet}
\label{sec:Appendix A}

%We now turn to the investigation of possible Hawking points in the CMB, as predicted by CCC. This is a type of problem, searching for a well-defined feature in a large and noisy data set, that is ideally suited for a machine learning approach. 

\begin{figure}[hbt!]
\centering
\includegraphics[scale=0.5]{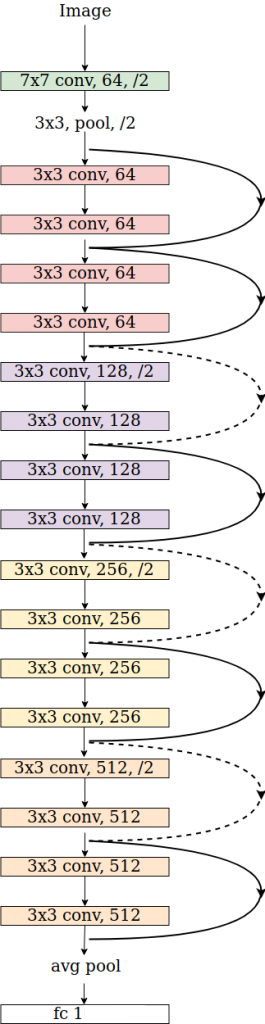}
\caption{The ResNet18 architecture used in HawkingNet. "Conv" and "Pool" labels refer to convolution and pooling layers. The numbers 3x3, 7x7, 64-512 represent the convolution kernel size, and the number of available filters. Fc 1 stands for "fully connected layer with one filter". The bold lines illustrate skipping connections discussed in the text, and dashed lines - skipping connection with the maximized pooling power. }
\label{fig:resnet}
\end{figure}

Our HawkingNet software \url{https://github.com/winger/hawking_net} is designed to search for Hawking points in the noisy large CMB data sets. Its architecture is fully based on the deep residual network (ResNet18) - a subclass of the convolution neural networks (CNN) family. ResNet is very popular in various scientific fields that demand visual grid-based image analysis. %: image and video classification, segmentation, and even recommendation systems used for predictive analysis. 
The HawkingNet consists of 18 neural layers - topological arrangements of convolutional (locally connected) neurons. The connectivity pattern is very much inspired by biological systems resembling the structure of the animal visual cortex, where the individual cortical neurons are stimulated only in a restricted region of the visual (receptive) field. The algorithm inputs the CMB data, processes the data through its internal network trained for data classification, and outputs the result in a form of a classification score, depending on how confident it is the image patch contains a Hawking point. The most important building blocks of the neural net are convolution layers and pooling layers (Figure ~\ref{fig:resnet}). 

\begin{figure}[hbt!]
\centering
\includegraphics[scale=0.8]{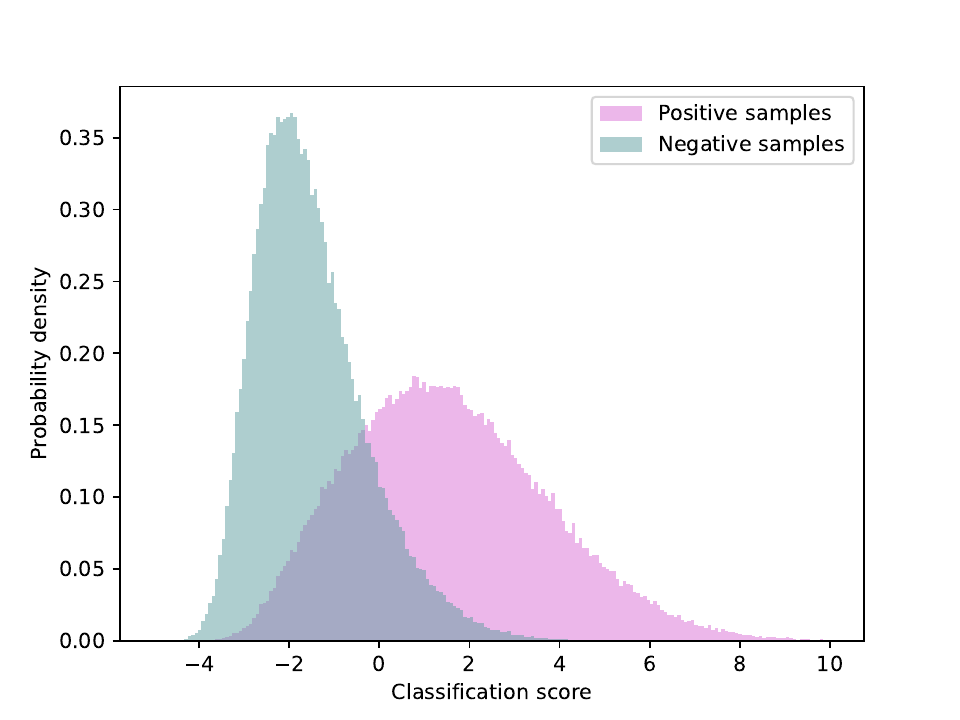}
\caption{The HawkingNet empirical probability distributions for the positive and negative samples. The positive samples indicate the data with classified Hawking points in contrast to the negative samples. }
\label{fig:ML_hist}
\end{figure}

First, the input CMB data goes through the convolution layers whose organization mimics the response of a neuron in the visual cortex to a specific stimulus in biological systems. Specifically, it performs the Frobenius inner product - a binary operation that takes the convolution kernel in a matrix form and the specific layer's input (also a matrix) and returns an output matrix. By convolving the input image in such a way, the number of initial free parameters is drastically reduced bypassing the input image deeper in the neural network a form of a abstract feature map. Next, the feature map is processed by the pooling layers which even more reduce the dimensions of data. These layers combine the outputs of neuron clusters in the feature map at one layer into a single neuron and pass it to the next layer. The uniqueness of the ResNet basis is the ability of the network to skip connections to jump over some layers, similarly to pyramidal cells in the cerebral cortex, where cortical layer VI neurons skipping intermediate layers on their way. Skipping the connection is beneficial both for the brain and for artificial neural net because it speeds up the learning process overall and mitigates larger training error for larger number of participating layers (known as Degradation problem). The final outcome from the HawkingNet performance is the %measured by ???
classification metric which analyses the feature map generated by the network. Classification in ML is generally a statistical model that uses a logistic function to model a binary dependent variable (the larger subgroup is referred to as logistic regression methods). So, in our case, the binary logistic function inputs the feature map from the residual network and defines the probability - a numerical value between 0 and 1 for every image patch whether the Hawking point is found or not in a form of positive and negative samples respectively (Figure~\ref{fig:ML_hist}). The network calculates the probability density distribution to account for the probability of finding the Hawking point in a considered data patch as a function of assigned classified scores. We see in Figure~\ref{fig:ML_hist} that most of the positive samples are more likely contain the Hawking point for classification score 2, while for the negative samples it is more likely for the classification score -2. These values are a function of the overall neural network sensitivity with a binary classifier which is typically measured by comparing its logistic function with the theoretical model. 

\begin{figure}[hbt!]
\centering
\includegraphics[scale=0.8]{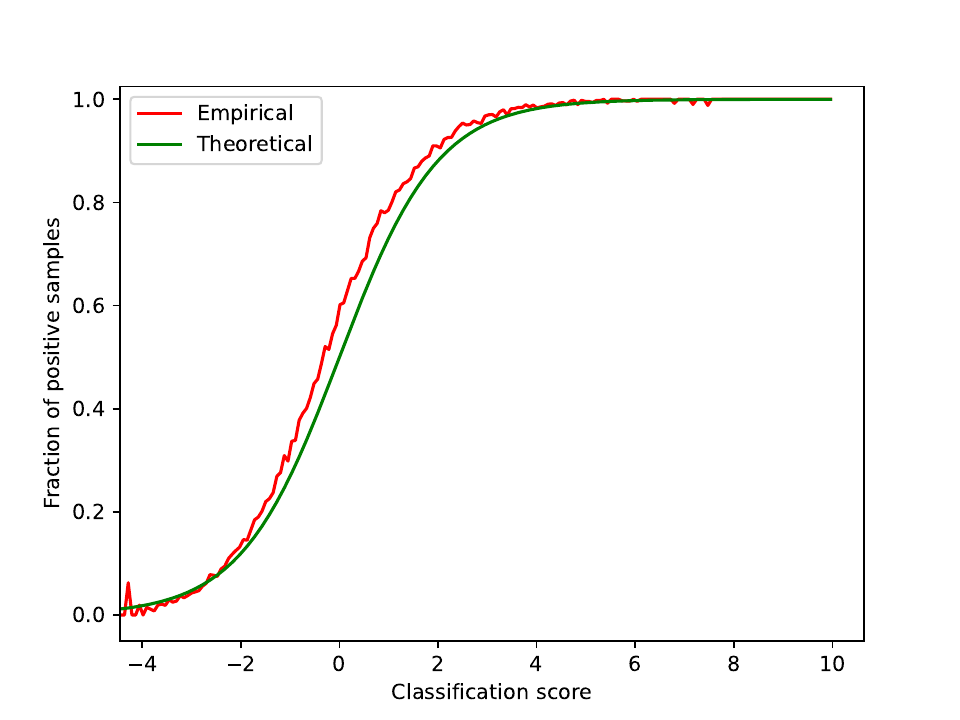}
\caption{The theoretical fraction of positive samples found by the HawkingNet- data with detected Hawking points as a function of the classification score. The aligned trend of both theoretical and empirical curves indicates maximized sensitivity of the neural network to the Hawking points.}
\label{fig:ML_logistic}
\end{figure}

Figure ~\ref{fig:ML_logistic} illustrates the measure of our sensitivity towards classifying Hawking points. On the y-axis, we show the theoretical fraction of positive samples with a given score, and the same fraction observed in practice at Figure ~\ref{fig:ML_hist}. Additionally, the aligned trends of empirical and theoretical curves indicate that our network sensitivity is maximized, which was done by 11 hours of neural network training.

\section{Planck COMMANDER spots of unusual temperature}
\label{sec:Appendix B}
\vspace{-.2in}
\begin{figure}[!hbt]
     \centering
     \begin{subfigure}[b]{0.45\textwidth}
         \centering
         \includegraphics[width=\textwidth]{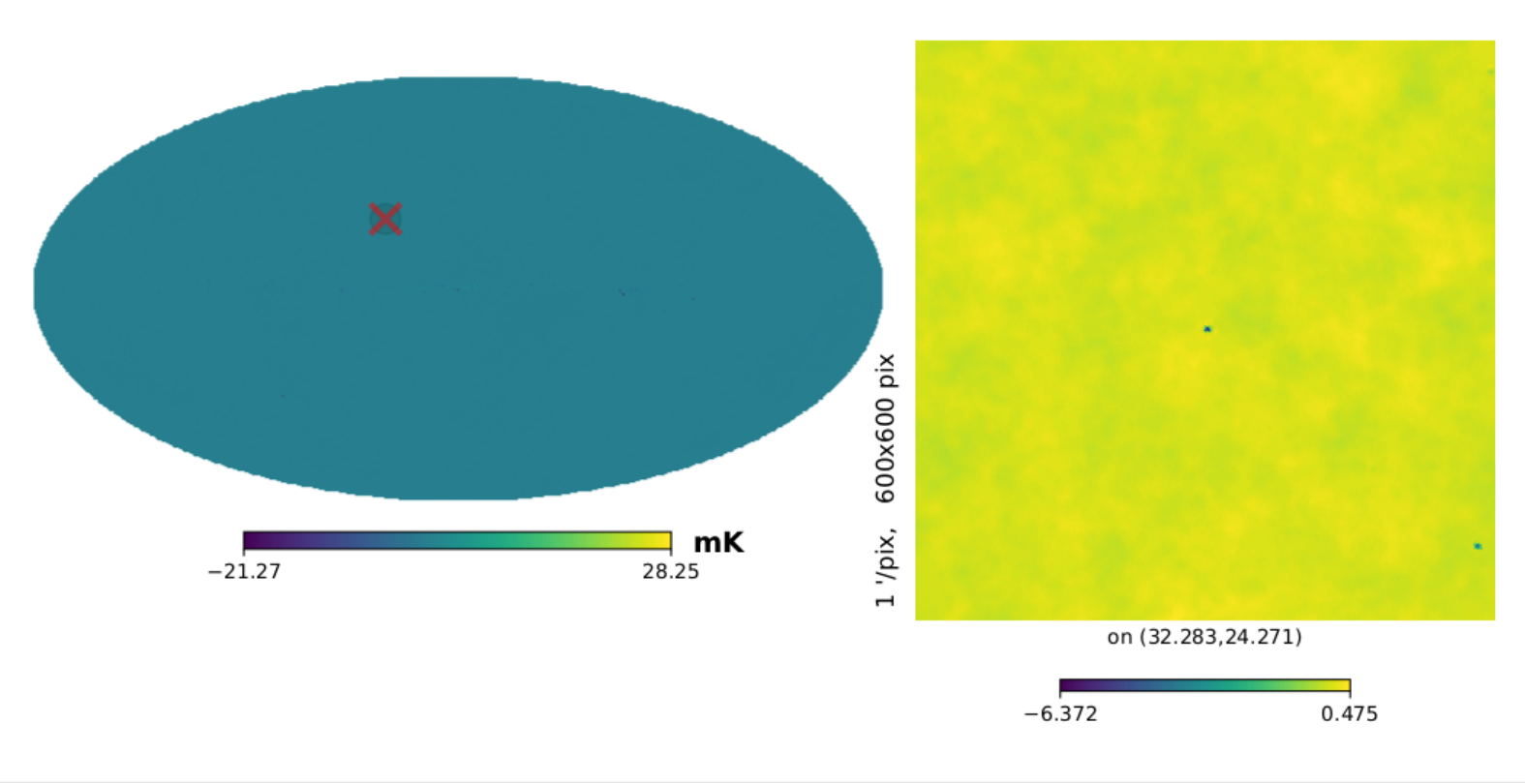}
         \caption{}
         \label{fig:fig21}
     \end{subfigure}
     \begin{subfigure}[b]{0.45\textwidth}
         \centering
         \includegraphics[width=\textwidth]{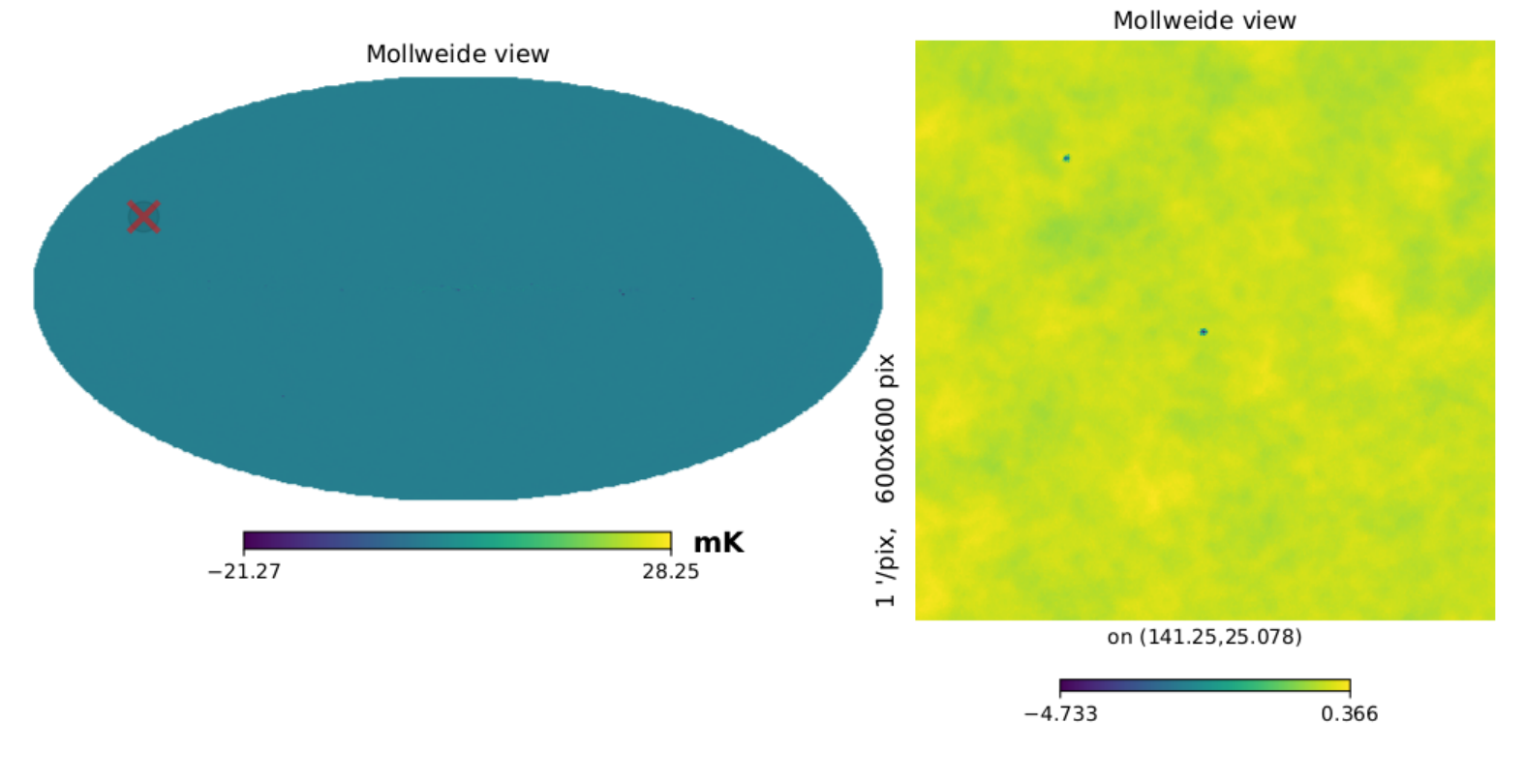}
         \caption{}
         \label{fig:fig22}
     \end{subfigure}
     \begin{subfigure}[b]{0.45\textwidth}
         \centering
         \includegraphics[width=\textwidth]{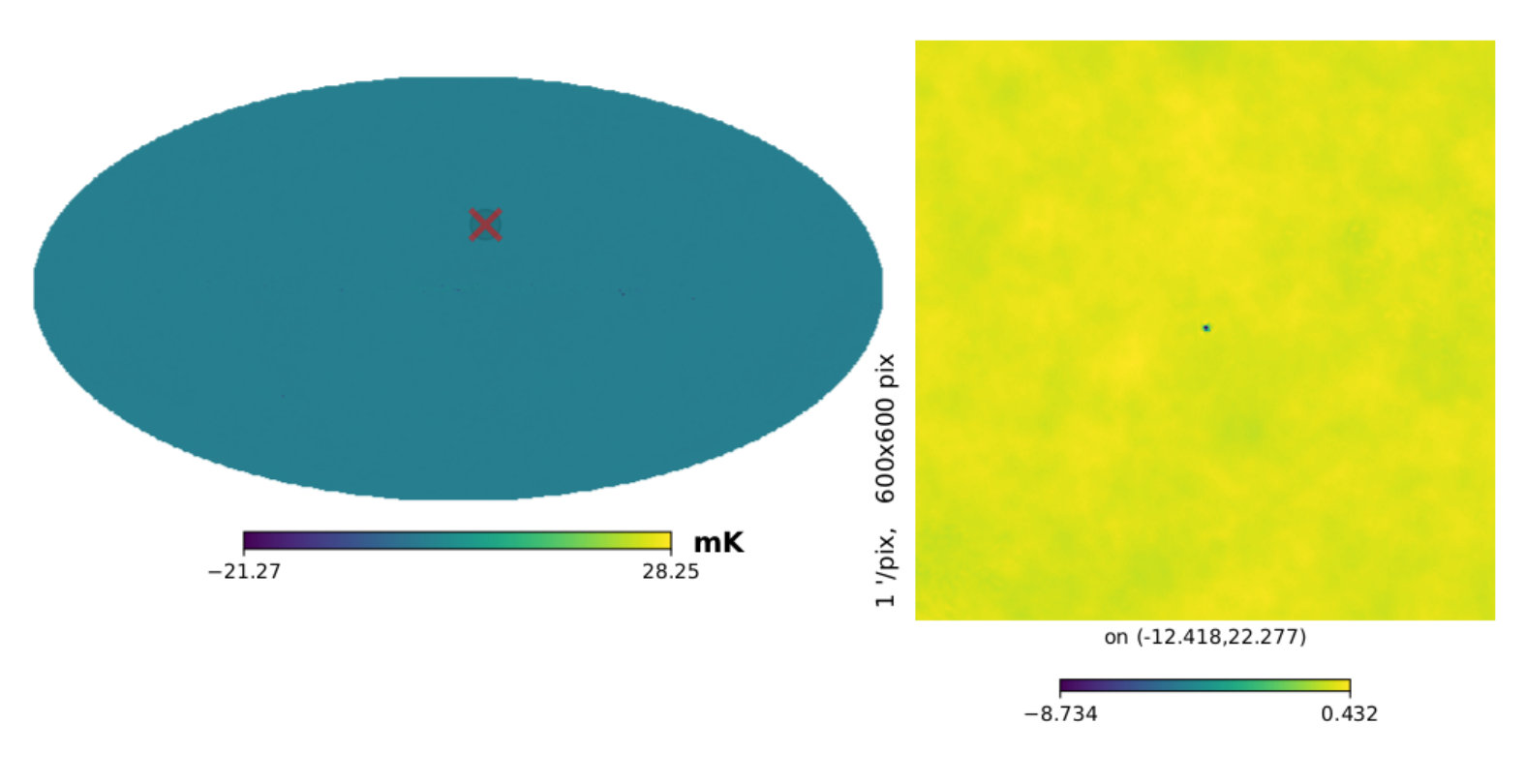}
         \caption{}
         \label{fig:fig23}
     \end{subfigure}
     \centering
     \begin{subfigure}[b]{0.45\textwidth}
         \centering
         \includegraphics[width=\textwidth]{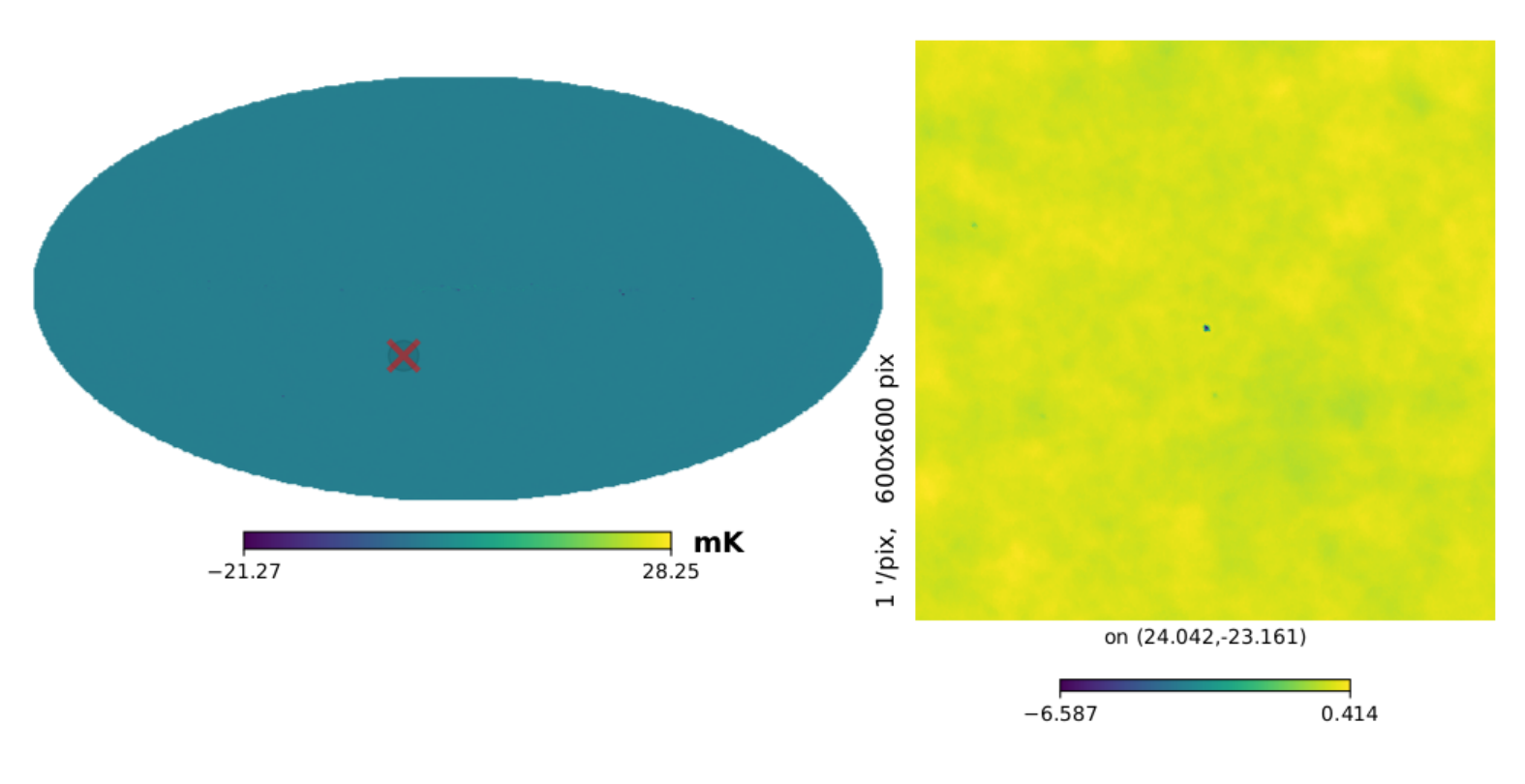}
         \caption{}
         \label{fig:fig24}
     \end{subfigure}
     \begin{subfigure}[b]{0.45\textwidth}
         \centering
         \includegraphics[width=\textwidth]{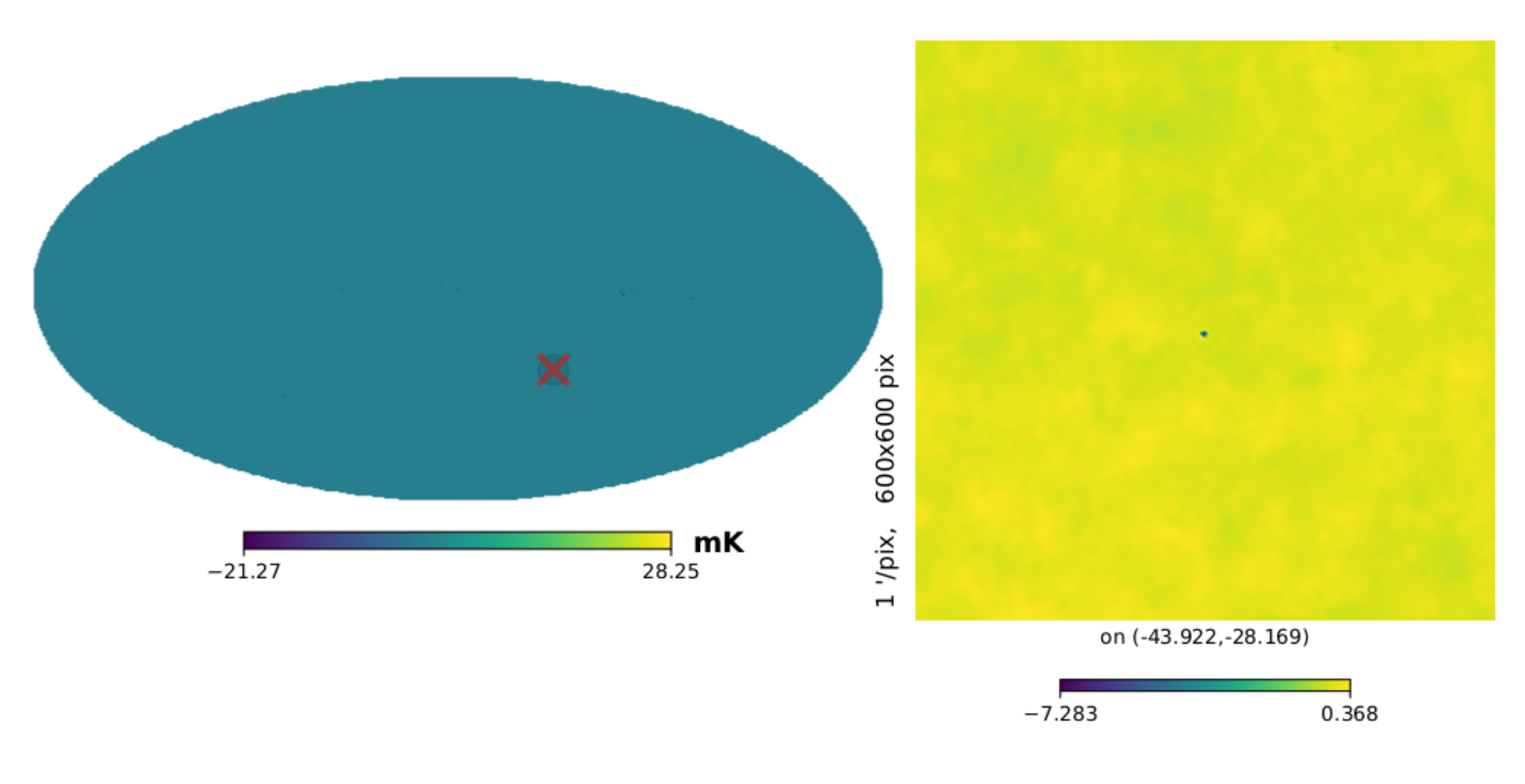}
         \caption{}
         \label{fig:fig25}
     \end{subfigure}
     \begin{subfigure}[b]{0.45\textwidth}
         \centering
         \includegraphics[width=\textwidth]{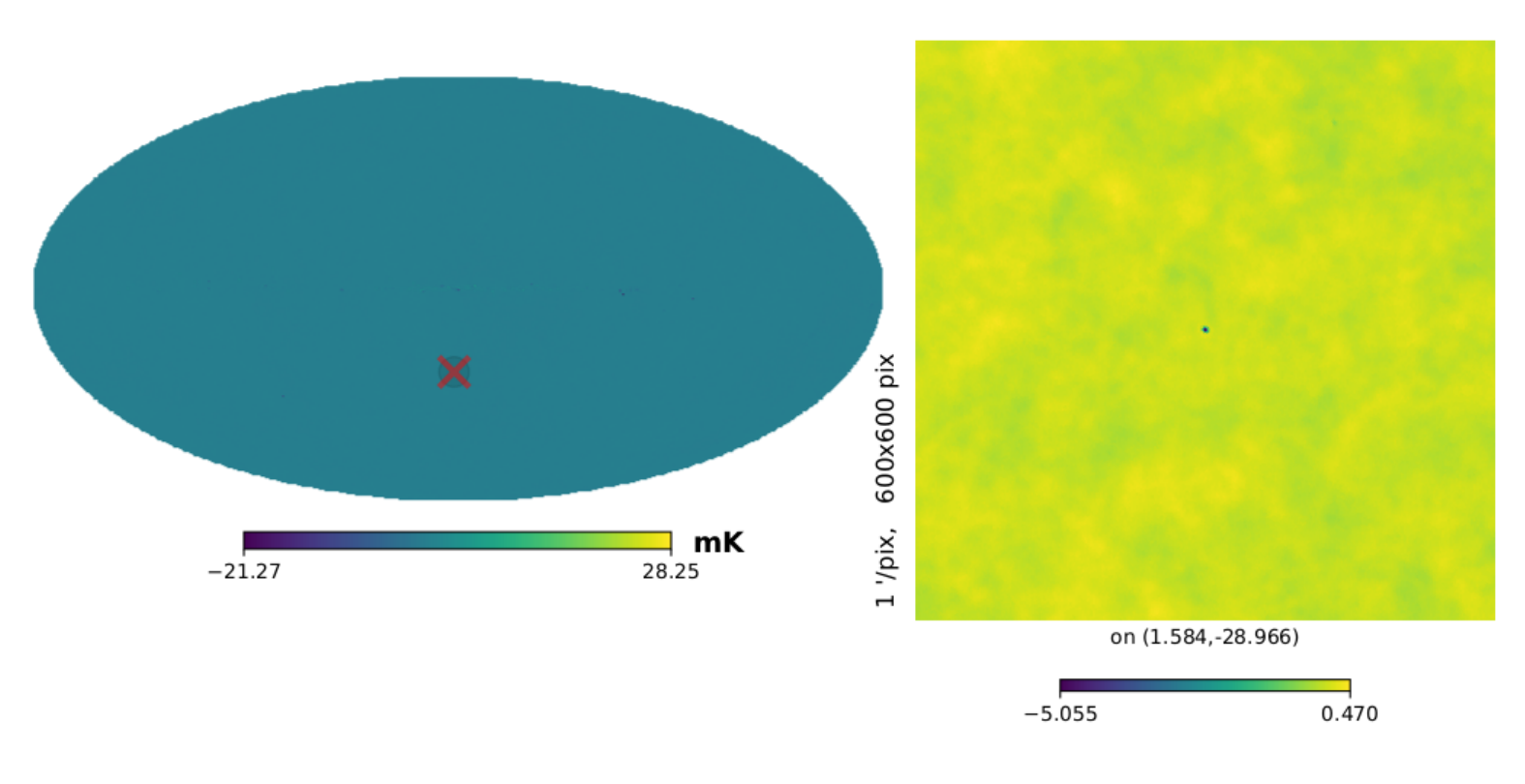}
         \caption{}
         \label{fig:fig26}
     \end{subfigure}
     \begin{subfigure}[b]{0.45\textwidth}
         \centering
         \includegraphics[width=\textwidth]{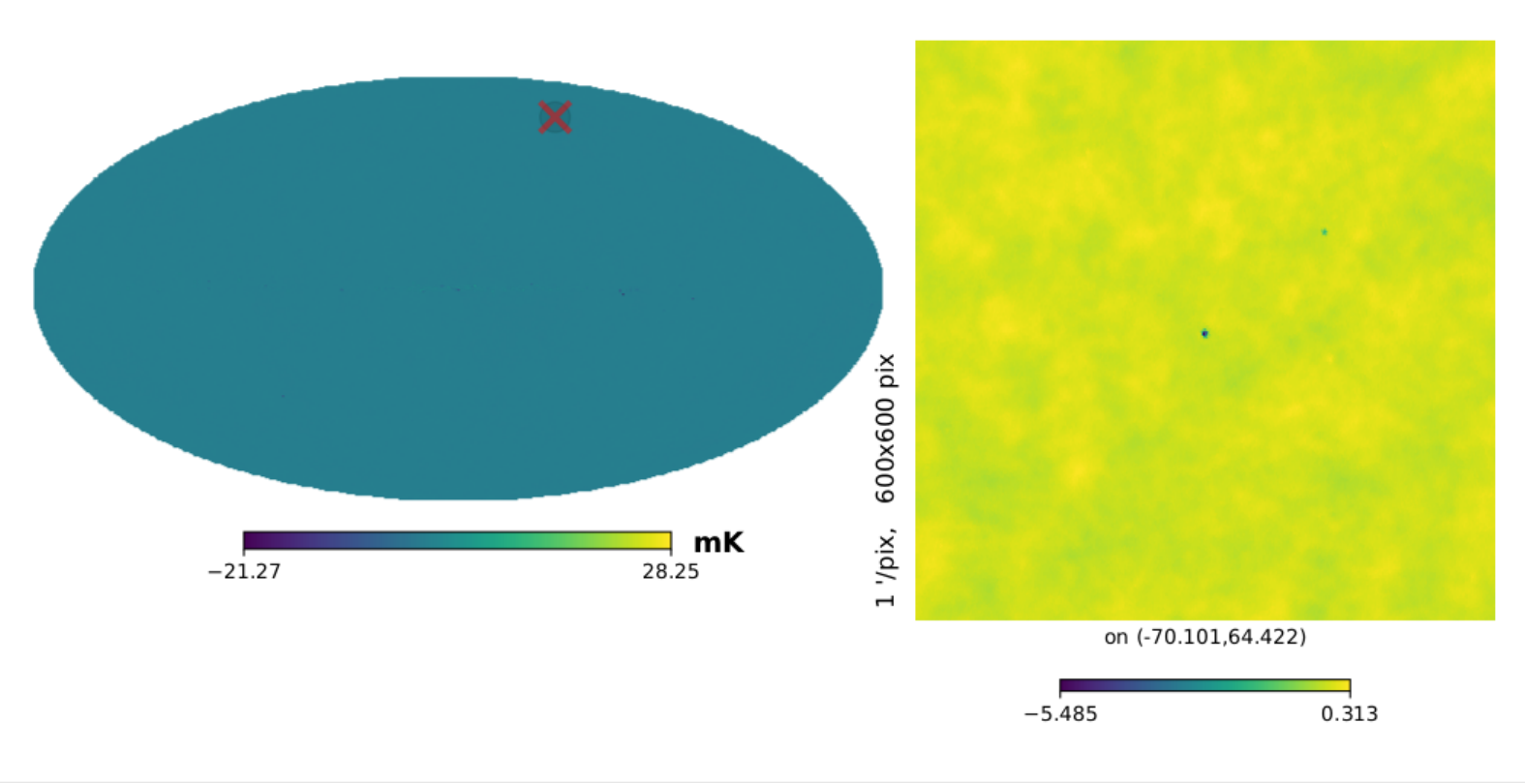}
         \caption{}
         \label{fig:fig20}
     \end{subfigure}
     \begin{subfigure}[b]{0.45\textwidth}
         \centering
         \includegraphics[width=\textwidth]{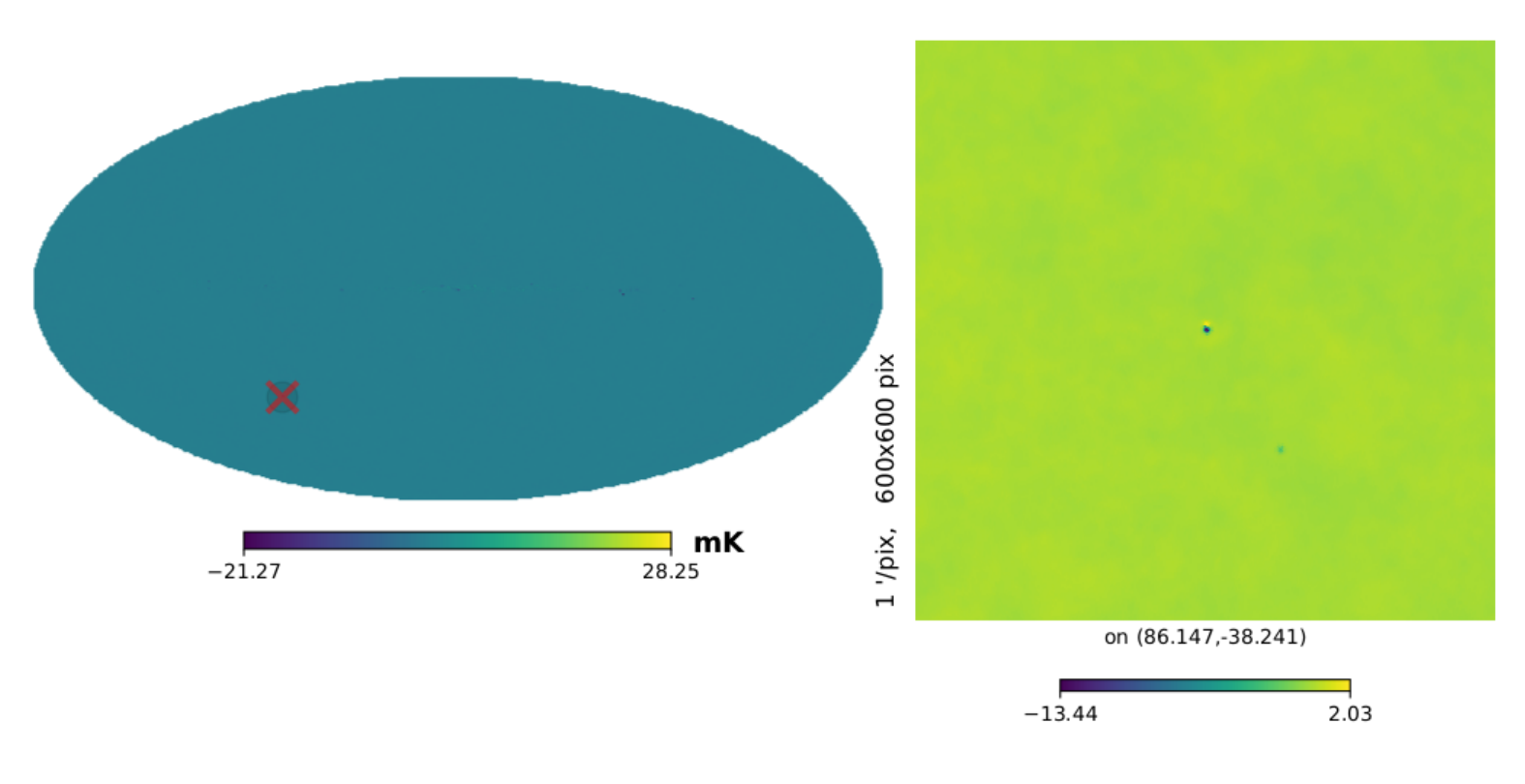}
         \caption{}
         \label{fig:fig28}
     \end{subfigure}
     \begin{subfigure}[b]{0.45\textwidth}
         \centering
         \includegraphics[width=\textwidth]{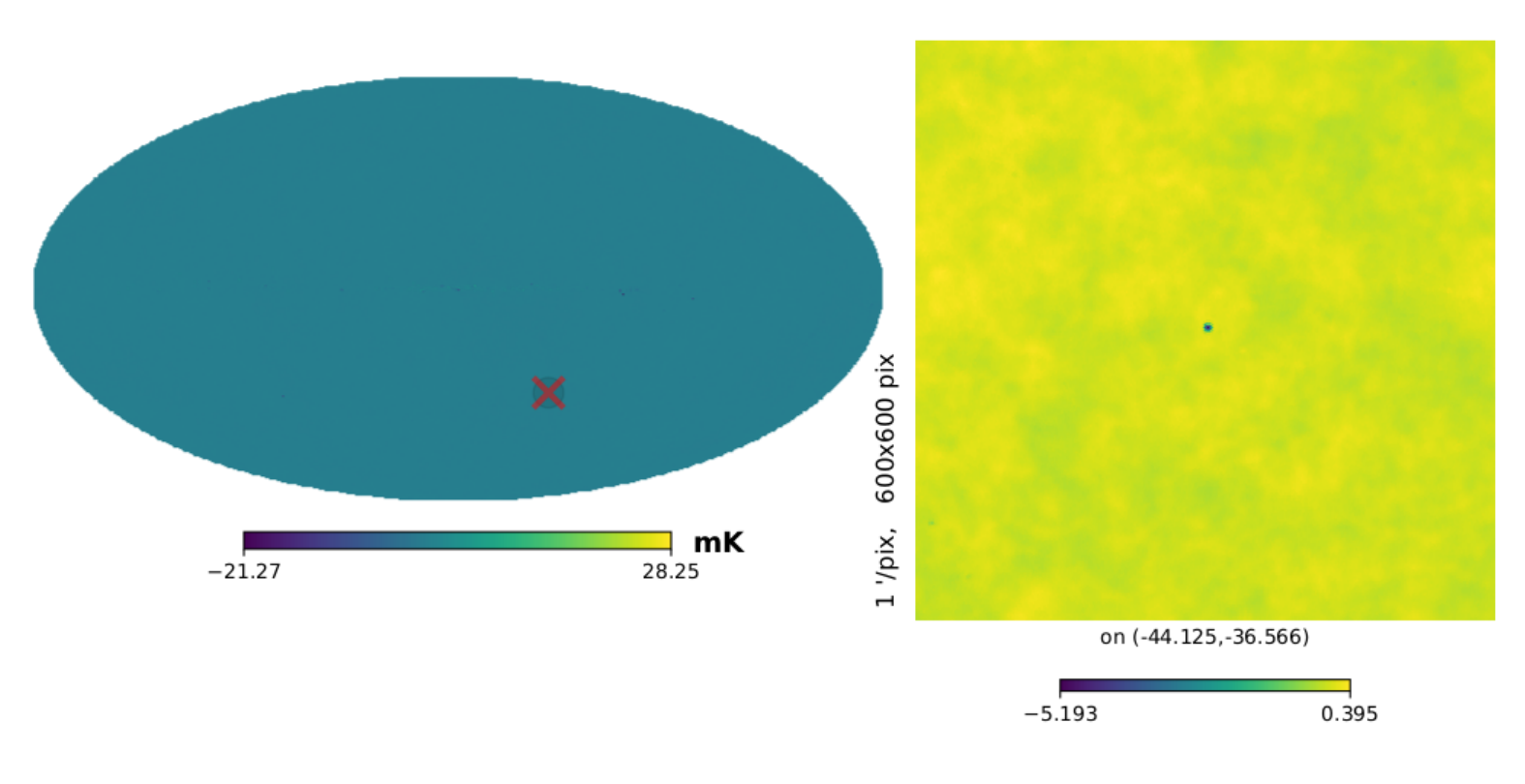}
         \caption{}
         \label{fig:fig29}
     \end{subfigure}
     \begin{subfigure}[b]{0.45\textwidth}
         \centering
         \includegraphics[width=\textwidth]{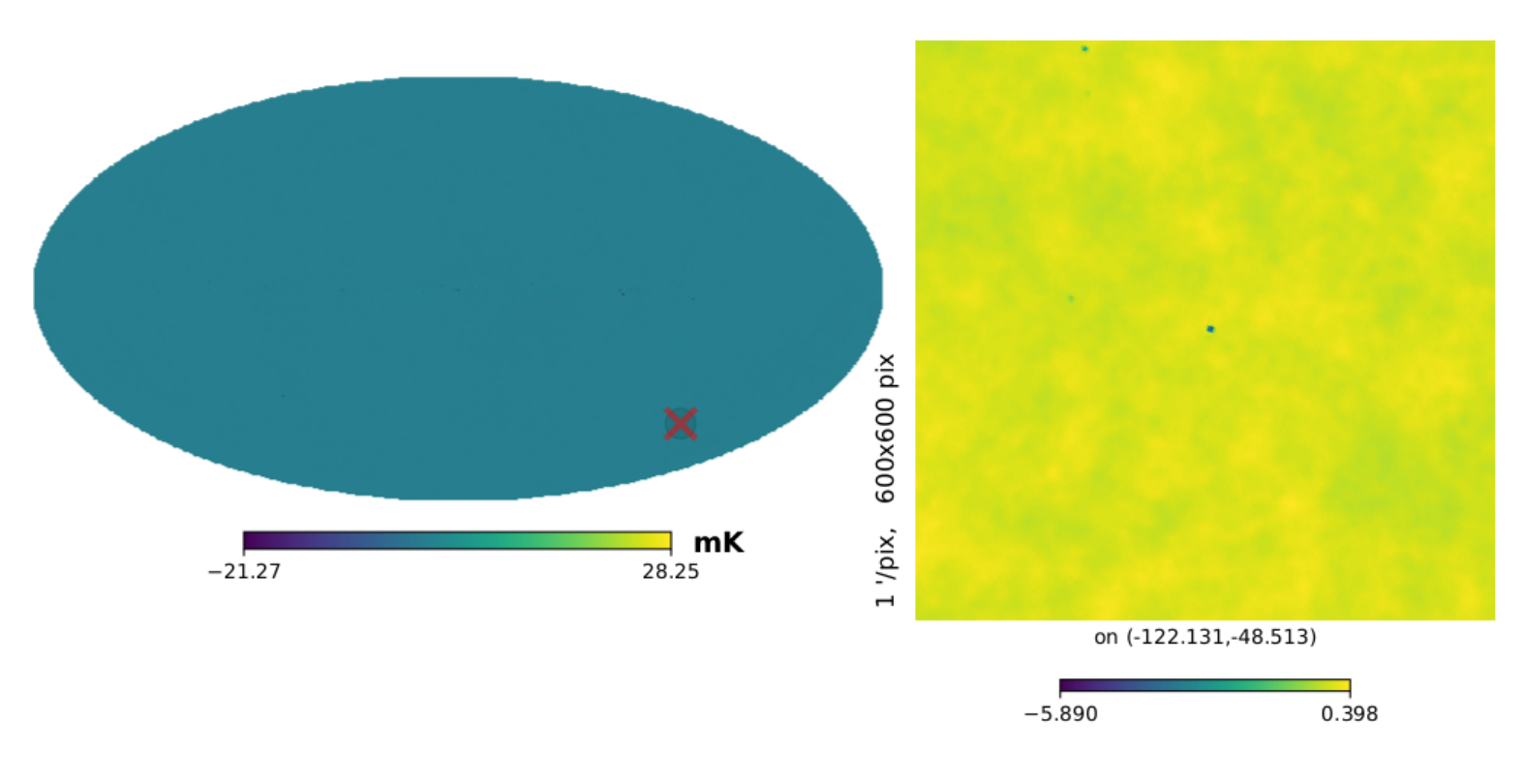}
         \caption{}
         \label{fig:fig30}
     \end{subfigure}
     \vspace{-.1in}
     \caption{ }
\label{fig:figureApendix}
\end{figure}

% \bibliography{apssamp}% Produces the bibliography via BibTeX.

\end{document}